\documentclass[%
prfluids,
reprint,
onecolumn,
superscriptaddress,
aps
]{revtex4-2}
\usepackage{graphicx}
\usepackage{dcolumn}
\usepackage{bm}
\usepackage{float}
\usepackage{amsmath,amssymb,amsfonts}
\usepackage{todonotes}
\usepackage{epstopdf}
\usepackage{diagbox}
\usepackage{tikz}

\usepackage{multirow}
\def\Xint#1{\mathchoice
	{\XXint\displaystyle\textstyle{#1}}%
	{\XXint\textstyle\scriptstyle{#1}}%
	{\XXint\scriptstyle\scriptscriptstyle{#1}}%
	{\XXint\scriptscriptstyle\scriptscriptstyle{#1}}%
	\!\int}

 \def\XXint#1#2#3{{\setbox0=\hbox{$#1{#2#3}{\int}$}
\vcenter{\hbox{$#2#3$}}\kern-.5\wd0}}

\def\XXiint#1#2#3{{\setbox0=\hbox{$#1{#2{\mathrm{#3\!\! #3}}}{\iint}$}
\vcenter{\hbox{$#2 {\mathrm{#3\!\! #3}}$}}\kern-.5\wd0}}
\DeclareMathOperator*{\argmin}{argmin}

\usepackage[caption=false]{subfig} 
\usepackage{ragged2e} 
\DeclareCaptionJustification{justified}{\justifying}

\usepackage{color}  
\definecolor{darkGreen}{rgb}{0,0.45,0}
\definecolor{darkBlue}{rgb}{0,0,0.7}
\definecolor{darkRed}{rgb}{0.76, 0.13, 0.28}
\definecolor{darkPurple}{rgb}{.5, 0, 1}
\definecolor{darkOrange}{rgb}{0.8, 0.33, 0.0}
\usepackage{hyperref}
\hypersetup{
	colorlinks=true, 
	linktoc=all,    
	linkcolor=darkBlue, 
	citecolor=darkGreen,
	urlcolor = darkRed,
}
\usepackage{comment}

\renewcommand{\bf}[1]{\mathbf{#1}}
\renewcommand{\d}{{\mathrm{d}}}
\newcommand{\degre}{\ensuremath{^\circ}}

\usepackage[margin=2.3cm]{geometry}
\usepackage{pifont}
\newcommand{\cmark}{\ding{51}}%
\newcommand{\xmark}{\ding{55}}%

\begin{document}

\title{Sail dynamics during tacking maneuvers}

\author{Christiana Mavroyiakoumou}%
\email[Electronic address: ]{cm4291@cims.nyu.edu}
\affiliation{Courant Institute, Applied Math Lab, New York University, New York, NY 10012, USA}

\author{Silas Alben}
\email[Electronic address: ]{alben@umich.edu}
\affiliation{Department of Mathematics, University of Michigan, Ann Arbor, MI 48109, USA}

\date{\today}

\begin{abstract}
We study the dynamics of sail membranes during a tacking maneuver, when the sail angle of attack is reversed in order to sail upwind. In successful tacking the sail flips to the mirror-image shape, while in unsuccessful tacking the sail remains stuck in a metastable state close to the initial shape.
We investigate whether the sail flips, and if so, how long it takes and the subsequent dynamics, over a parameter space that describes the sail membrane properties and the kinematics of tacking. We find that the ``steady" parameters---stretching rigidity, pretension, and final angle of attack---mostly determine whether a membrane flips or not. Flipping is more likely with larger values of the stretching rigidity, pretension, and final angle of attack. The dynamical parameters---membrane mass density, concavity of angle-of-attack transition kinematics, and time-length of the tacking maneuver---mainly affect how long flipping takes. With large membrane mass the membrane can maintain the momentum from the tacking motion long enough to reach the flipped state, but it may also take longer to converge to a steady shape. Concave-down angle-of-attack profiles and small tack times are generally associated with shorter flip times but a few exceptions exist, because these kinematics can also give a larger acceleration to the fluid-sail system that persists for longer times. We also investigate slack sails and find that they are more difficult to flip.



\end{abstract}


\maketitle

\section{Introduction}

Membranes are thin, extensible sheets with negligible bending modulus that are stable in a variety of configurations, and are thus widely used for example, in sails~\cite{nielsen1963theory,newman1984two,newman1987aerodynamic,smith1995computation,vanden1981shape,vanden1982nonlinear,kimball2009physics}, parachutes~\cite{pepper1971aerodynamic,stein2000parachute}, micro-air vehicles~\cite{shyy1999flapping,lian2003membrane,albertani2007aerodynamic,hu2008flexible,stanford2008fixed}, ballutes for space exploration~\cite{scott2007aeroelastic,rohrschneider2007survey}, 
supersonic aircraft and rockets~\cite{voss1961effect,ashley1956piston},
roofs in civil engineering~\cite{haruo1975flutter,knudson1991recent,sygulski1996dynamic,sygulski1997numerical,sygulski2007stability}, and naturally in the wings of flying animals~\cite{swartz1996mechanical,song2008aeromechanics,cheney2015wrinkle}. 
In this work we study membrane motions inspired by sail dynamics. The sail is a thin membrane wing that generates propulsive forces in a high-Reynolds-number (nearly inviscid) fluid flow. The sail is given a particular shape and direction depending on wind conditions to generate such forces~\cite{kimball2009physics,anderson2008physics}. 
Optimizing the sail membrane shape and position involves the adjustment of various controls attached to the sail. The majority of racing yachts have three types of sails: a mainsail, a headsail, and a spinnaker~\cite{maria2013recent,kimball2009physics,anderson2008physics,lombardi2012strongly}. Here we focus only on the mainsail.

\begin{figure}[H]
    \centering
\includegraphics[width=\textwidth]{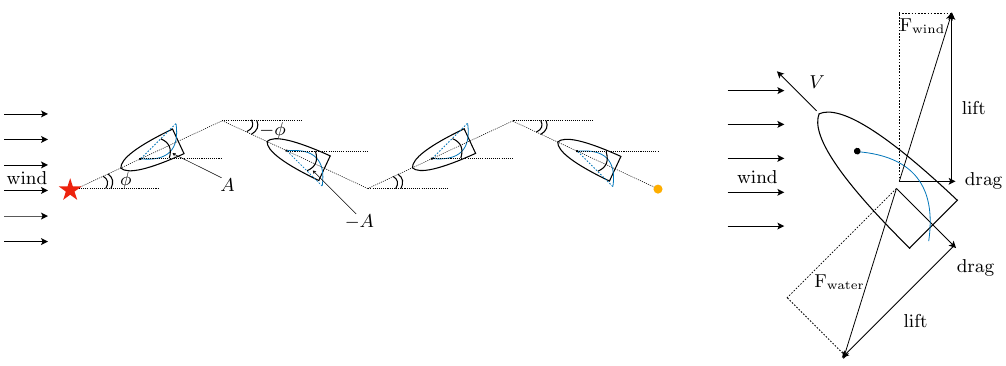}
    \caption{Schematic diagram showing on the left-hand side a series of three tacking maneuvers, from the starting point (orange circle) to the final destination (red star) giving a zig-zag path of the sailboat. The sail membrane is shown as a blue solid curve and the chord connecting the membrane's leading and trailing edges is shown as a blue dashed line. $\phi$ is the angle that the sailboat makes with the wind and $A$ is the angle that the sail's chord makes with the horizontal. On the right-hand side we show a force diagram illustrating the lift and drag components of the wind and water forces on a sailboat. When the sailboat is moving at a constant speed $V$, $\mathrm{F}_{\mathrm{wind}}$ is equal and opposite to $\mathrm{F}_{\mathrm{water}}$. The force diagram is an adapted version of~\cite[fig.~3.4]{kimball2009physics}. 
    }
    \label{fig:schematic}
\end{figure}

Specifically we consider a maneuver for upwind sailing known as \textit{tacking}, shown schematically on the left-hand side of figure~\ref{fig:schematic}. The sailboats perform a kind of zig-zag motion that consists of a series of consecutive tacking maneuvers, changing their angle to the wind from $-\phi$ to $\phi$ and vice versa. Fore-and-aft sails generate a significant force at right angles to the apparent wind, a force that pushes the sailboat obliquely upwind, alternately to either side of the wind direction during the tacking maneuver. This way the sailboat can move towards its final destination (red star), which is in the upwind direction. This maneuver allows the sailboat's forward direction to rotate through a range of angles (the so-called no-sail zone), so that the direction from which the wind blows changes from one side of the boat to the other, allowing the sailboat to progress in the desired target direction. The no-sail zone is a specific range of boat velocity directions relative to the wind, centered on the upwind direction, for which the sail cannot propel the boat forward because the lift force does not have a positive component in the boat's forward direction~\cite{kimball2009physics,petres2012potential}. If instead the boat alternately moves in directions on either side of the no-sail zone, its net motion can be upwind.

When membranes are held with their ends fixed in a uniform oncoming fluid flow, they tend to adopt steady (or nearly steady) shapes that curve to one side~\cite{song2008aeromechanics,mavroyiakoumou2020large}.
The sail (blue curve) with an angle of attack (the angle from the horizontal to its chord) labeled $-A$ in figure~\ref{fig:schematic} is initially cambered (i.e.\ curved) to one side.
After a successful tacking it flips to the mirror-image configuration (closest to the red star) with angle of attack $A$.

Sailing upwind is made possible by the interplay of forces generated by both the wind acting on the sail and the water that acts on the hull and the keel (a plate that extends downward from the bottom of the hull along its centerline). A diagram of the force components acting on the sailboat is shown on the right-hand side of figure~\ref{fig:schematic}. 
During sailing, there are drag forces due to the wind and water that are in the direction opposite to the boat's velocity relative to the wind and water, respectively. The lift forces are perpendicular to the drag forces.
The sums of the lift and drag forces from the wind and the water are denoted by $\mathrm{F}_{\mathrm{wind}}$
and $\mathrm{F}_{\mathrm{water}}$, respectively.
Upwind sailing is possible if $\mathrm{F}_{\mathrm{wind}}$ is positively correlated with the direction of the boat so that thrust is generated to balance drag forces on the boat.


During the tacking maneuver, lift decreases and the fluid forces are generally opposite to the boat's velocity, so the boat speed decreases. So it is important that the maneuver is performed only when the boat's momentum is sufficient to keep it moving forward through the turn. Naturally, because of this reduction in the sailboat speed, after the tacking maneuver (or ``tack") the sailboat must start regaining speed.
Therefore the best tack is not necessarily the fastest tack; it is the tack in which the least speed is lost during the maneuver~\cite{kimball2009physics}.

Sail membranes are required to generate large lift forces with small drag. The sail's sharp leading edge promotes flow separation but its membrane structure offers high flexibility that allows it to adapt to the unsteady airflow conditions quickly and lead to reduction of separation. In~\cite{souppez2019recent,arredondo2023vortex} they focus on understanding the underlying force generation mechanisms of sails and on the force contribution of vortical flow structures in separated flow regions. Newman in~\cite{newman1987aerodynamic} mentions that unlike rigid wings, sail membranes must also avoid a region of angles of attack near zero where 
the sail begins to lose its single-hump shape (e.g.\ figure~\ref{fig:schematic}), becoming S-shaped and possibly oscillating (i.e.\ ``luffing")~\cite{greenhalgh1984aerodynamic}. Remaining in this S-shape or exhibiting a local flutter or full-sail flutter instability~\cite{hess1961experimental} is undesirable. Discussions of sails remaining in the S-shaped state can also be found on sailing websites~\cite{windsportatlanta,cruisersforum,catsailor}.

In the sail tacking maneuver that we consider, the sail is induced to move from one local equilibrium to another. A similar transition of a bistable flexible structure in an oncoming flow---a buckled elastic sheet with two clamped edges---was considered by~\cite{kim2021flow,kim2021snap} experimentally and by~\cite{chen2023snap} numerically using the penalty immersed boundary method. 
These studies considered buckled elastic sheets for energy harvesting applications. \cite{chen2023snap}~states that higher output voltages and greater energy harvesting can be achieved than for a related system, an inverted flag~\cite{shoele2016energy,orrego2017harvesting,kim2013flapping,ryu2015flapping,gurugubelli2015self}. \cite{gomez2017passive} studied snap-through of a sheet by fluid-dynamic
loading at very low Reynolds number~$\mathcal{O}(10^{-2})$.

In recent years, advances in automated vehicle technology have created opportunities for smart urban mobility~\cite{jouffroy2009control} and ocean exploration~\cite{sanchez2020autonomous,bandyopadhyay2005trends}. In \cite{tranzatto2015debut,tranzatto2015navigation} they consider different tack maneuvers for autonomous sailboats~\cite{petres2012potential,stelzer2010reactive,xiao2012wind} and deploy a cost function method that takes into account the progress towards the target, the presence of obstacles, sailing constraints such as the no-sail zone, the time spent in maneuvering, and other tactical considerations. This cost function is evaluated in real time based on the available measurements. 
A~review of sail aerodynamics is given in~\cite{maria2013recent} and an overview along with control strategies of autonomous sailboats can be found in~\cite{tipsuwan2023overview}. For a real racing sailboat, \cite{hansen2019maneuver} optimized the tacking maneuver as well as other sailboat controls to minimize the distance lost during tacking.

Many previous studies considered steady sail configurations. Some studies also considered unsteady oscillations about the steady shape~\cite{waldman2013shape,sygulski2007stability}.
In this paper we consider a conventional tacking maneuver but a few other unsteady sail motions have also been analyzed previously.
\cite{gerhardt2011unsteady} examined instead an unsteady sail motion---a sinusoidal oscillation that models the effect of water waves on the boat---and found that very little energy can be extracted from the unsteady flow about the sails. Specifically the time-varying components of the aerodynamic forces are small and the thrust gain is minimal. \cite{young2019effect} also considered sinusoidal oscillations of a rigid model sail in both the transverse and streamwise directions to investigate the possibility of increasing thrust, a maneuver called ``sail flicking''. Related maneuvers for increased thrust are considered in~\cite{schutt2016unsteady}.

The structure of this paper is as follows. In \S\ref{sec:model}, we present the membrane and vortex sheet model. In \S\ref{sec:steady} we show the steady membrane behavior at different values of steady angles of attack, i.e. where the model does not involve the tacking maneuver, whereas in \S\ref{sec:tacking} we show examples of the dynamics after a tack is performed. In \S\S\ref{sec:membraneParameters}--\ref{sec:kinematicParameters} we study in detail the effects of seven different parameters on the sail membrane dynamics when the tacking maneuver is performed, divided into two main parts: those that determine the properties of the sail membrane and those that control the kinematics of the tack. In \S\ref{sec:slack} we consider an alternative configuration---a ``slack" sail, where the membrane ends are brought together, reducing the pretension before the tacking maneuver. \S\ref{sec:conclusions} presents the conclusions.




\section{Membrane and vortex-sheet model}\label{sec:model}

\begin{figure}[H]
    \centering
    \includegraphics[width=0.6\linewidth]{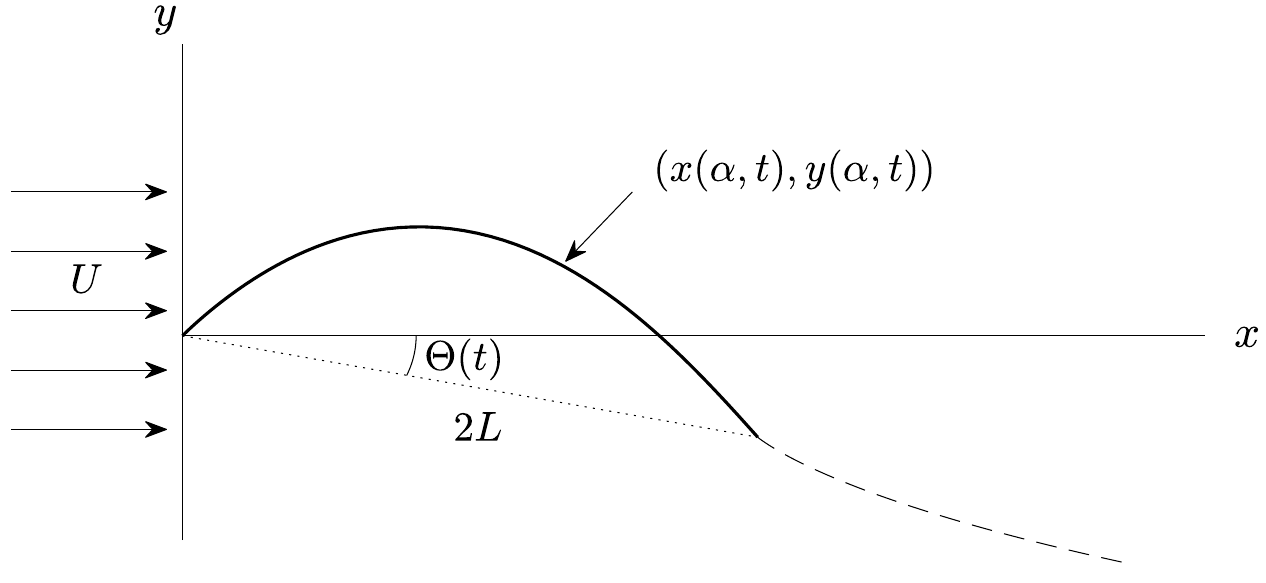}
    \caption{Schematic diagram of a flexible sail membrane (solid curved line) at an instant in time. Here $2L$ is the chord length (the distance between the endpoints), $U$ is the wind velocity relative to the leading edge, $\Theta(t)$ is the instantaneous angle of attack,  $(x(\alpha,t),y(\alpha,t))$ is the membrane position, and the dashed line is the free vortex sheet wake.}
    \label{fig:schematicMembrane}
\end{figure}

Our previous work~\cite{mavroyiakoumou2020large,mavroyiakoumou2021dynamics} considered passive membrane flutter dynamics induced by a small perturbation from the zero angle-of-attack state. Instead, in the current work we consider membranes whose trailing edge is being moved with specified kinematics with nonzero angles of attack. We study the resulting motions with the membrane and flow equations used in~\cite{mavroyiakoumou2020large}, which we repeat briefly here for completeness, together with the new time-dependent boundary conditions we consider. We start with a membrane that is nearly aligned with a two-dimensional background fluid flow that has speed $U$ in the far field (shown schematically in figure~\ref{fig:schematicMembrane}). We bring the membrane's trailing edge from an angle of attack $-A$ to the desired final angle of attack $A$. We reiterate that the angle of attack is defined as the angle between the sail chord and the apparent wind direction (the angle shown in the schematic diagram on the left-hand side of figure~\ref{fig:schematic}). The instantaneous angle of attack is denoted by $\Theta(t)$, and reaches $A$ after the tack.

The membrane dynamics are described by the unsteady extensible membrane equation with body inertia, stretching resistance, and fluid pressure loading, obtained by writing a force balance equation for a small section of membrane that lies between material coordinates $\alpha$ and $\alpha+\Delta \alpha$:
\begin{equation}\label{eq:dimensionalMembrane}
    \rho_s hW \partial_{tt}\zeta(\alpha,t)\Delta \alpha=T(\alpha+\Delta\alpha)\mathbf{\hat{s}}-T(\alpha,t)\mathbf{\hat{s}}-[p]_-^+(\alpha,t)\mathbf{\hat{n}}W(s(\alpha+\Delta\alpha,t)-s(\alpha,t)).
\end{equation}
Here $\rho_s$ represents the mass per unit volume of the undeflected membrane, $h$ is the membrane's thickness, and~$W$ its spanwise width, all uniform along the length. Like the fluid flow, the motions of the sail membrane are assumed to be invariant in the spanwise direction (along~$W$), and the effect of gravity is neglected for simplicity. In~\eqref{eq:dimensionalMembrane},
$\zeta(\alpha,t) = x(\alpha,t)+iy(\alpha,t)$ denotes the membrane position in the complex plane, parameterized by the material coordinate $\alpha$, $-L\leq \alpha\leq L$ ($L$ is half the chord length, or the initial chord length when it varies with time, at the end of paper) and $t$ is time. $T$ is the tension in the membrane, $[p]_-^+$ is the pressure jump across it, $s(\alpha,t)$ is the local arc length coordinate, and the unit vectors tangent and normal to the membrane are $\mathbf{\hat{s}}=\partial_\alpha\zeta(\alpha,t)/\partial_\alpha s(\alpha,t)=e^{i\theta(\alpha,t)}$ and $\mathbf{\hat{n}}=i\mathbf{\hat{s}}=ie^{i\theta(\alpha,t)}$, respectively, with $\theta(\alpha,t)$ the local tangent angle and $\partial_\alpha s$ the local stretching factor.
We use~$+$ to denote the side towards which the membrane normal~$\mathbf{\hat{n}}$ is directed, and~$-$ for the other side. However, for the remainder of this paper, we usually drop the $+$ and $-$ for ease of notation. 

Dividing~\eqref{eq:dimensionalMembrane} by $\Delta\alpha$ and taking the limit $\Delta\alpha\to 0$, we obtain
\begin{equation}\label{eq:dimensionalMembraneEq}
    \rho_shW\partial_{tt}\zeta(\alpha,t) = \partial_\alpha(T(\alpha,t)\mathbf{\hat{s}})-[p](\alpha,t)W\partial_\alpha s\mathbf{\hat{n}},
\end{equation}
where the membrane tension $T(\alpha,t)$ is given by linear elasticity~\cite{carrier1945non,narasimha1968non,nayfeh2008linear} as
\begin{equation}
    T(\alpha,t)=\overline{T}+EhW(\partial_{\alpha} s(\alpha,t)-1).
\end{equation}
Here $E$ is the Young's modulus and $\overline{T}$ is the tension in the initial, undeflected equilibrium state. After nondimensionalizing length by $L$, time by $L/U$, and pressure by $\rho_f U^2$, where $\rho_f$ is the density of the fluid and $U$ is the oncoming flow velocity, \eqref{eq:dimensionalMembraneEq} becomes the nonlinear, extensible membrane equation
\begin{equation}\label{eq:membrane}
R_1\partial_{tt}\zeta -\partial_\alpha ((T_0+R_3(\partial_\alpha s-1))\mathbf{\hat{s}})=-[p]\partial_\alpha s\mathbf{\hat{n}}.
\end{equation}
In \eqref{eq:membrane},  $R_1=\rho_sh/(\rho_f L)$ is the dimensionless membrane mass, $T_0=\overline{T}/(\rho_f U^2L W)$ is the dimensionless pretension, and finally, $R_3=Eh/(\rho_f U^2 L)$ is the dimensionless stretching rigidity. 

We express the 2D flow past the membrane using $z=x+iy$, the complex representation of the $xy$ flow plane. 
As in~\cite{mavroyiakoumou2020large}, the inviscid flow can be represented by a vortex sheet (a curve across which the tangential velocity component is discontinuous~\cite{saffman1992vortex}) whose position and strength evolve in time. The vortex sheet consists of a `bound' part and a `free' part. The bound vortex sheet coincides with the membrane for $-1\leq\alpha\leq 1$ and the free vortex sheet emanates from the trailing edge of the membrane at $\alpha=1$. They both have strength densities denoted by $\gamma$ and positions denoted by $\zeta$.
The complex conjugate of the fluid velocity at any point $z$ not on the vortex sheets is a sum of the horizontal background flow with speed unity and the flow induced by the bound and free vortex sheets,
\begin{equation}
u_x(z)-iu_y(z)= 1 +\frac{1}{2\pi i}\int_{-1}^1\frac{\gamma(\alpha,t)}{z-\zeta(\alpha,t)}\partial_\alpha s \,\d\alpha+\frac{1}{2\pi i}\int_0^{s_{\max}(t)} \frac{\gamma(s,t)}{z-\zeta(s,t)}\d s.
\end{equation}
To determine the bound vortex sheet strength $\gamma$ we require that the fluid does not penetrate the membrane, which is known as the kinematic boundary condition. Here $\gamma$ represents the jump in the component of the flow velocity tangent to the membrane from the $-$ to the $+$ side, i.e.,\ $\gamma=-[(u_x,u_y)]^+_-\cdot\mathbf{\hat{s}}$. The normal components of the fluid and membrane velocities are equal along the membrane:
\begin{equation}\label{eq:kinematic}
    \mathrm{Re}(\mathbf{\hat{n}}\partial_t\overline{\zeta}(\alpha,t)) =\mathrm{Re}\left\{ \mathbf{\hat{n}}\left( 1 +\frac{1}{2\pi i}\int_{-1}^1\frac{\gamma(\alpha,t)}{z-\zeta(\alpha,t)} \partial_\alpha s\,\d\alpha+\frac{1}{2\pi i}\int_0^{s_{\max}(t)} \frac{\gamma(s,t)}{z-\zeta(s,t)}\d s \right)\right\},
\end{equation}
where $\mathbf{\hat{n}}$ is written as a complex scalar. Solving~\eqref{eq:kinematic} for $\gamma$ requires an additional constraint that the total circulation is zero for a flow started from rest, by Kelvin's circulation theorem. At each instant the part of the circulation in the free sheet, or alternatively, the strength of $\gamma$ where the free sheet meets the trailing edge of the membrane, is set by the Kutta condition which makes velocity finite at the trailing edge. At every other point on the free sheet, $\gamma$ is set by the criterion that circulation (the integral of $\gamma$) is conserved at fluid material points of the free sheet. We evolve the position of the free vortex sheet using the Birkhoff-Rott equation as described in~\cite{mavroyiakoumou2020large}.

The vortex sheet strength $\gamma(\alpha,t)$ is coupled to the pressure jump $[p](\alpha,t)$ across the membrane using a version of the unsteady Bernoulli equation written at a fixed material point on the membrane:
\begin{equation}\label{eq:pressure}
    \partial_\alpha s\partial_t\gamma +\partial_\alpha\left(\gamma(\mu-\tau)\right)+\gamma(\partial_\alpha \tau-\nu\kappa\partial_\alpha s)=\partial_\alpha [p].
\end{equation}
In \eqref{eq:pressure}, $\kappa(\alpha,t)=\partial_\alpha\theta/\partial_\alpha s$ is the membrane's curvature, $\mu$ is the tangential component of the average flow velocity at the membrane,
\begin{equation}\label{eq:mu}
\mu(\alpha,t)=\text{Re}\left\{\hat{\mathbf{s}}\left( 1+\frac{1}{2\pi i} \Xint-_{-1}^{1} \frac{\gamma(\alpha',t)\partial_\alpha s(\alpha',t)}{\zeta(\alpha,t)-\zeta(\alpha',t)}\,\text{d}\alpha'+\frac{1}{2\pi i }\int_{0}^{s_{\max}(t)}\frac{\gamma(s,t)}{\zeta(\alpha,t)-\zeta(s,t)} \text{d} s\right)\right\},
\end{equation}
and $\tau$ and $\nu$ are the components of the membrane's velocity tangent and normal to itself,
respectively:
\begin{equation}\label{eq:tau}
\tau(\alpha,t)=\text{Re}\left(\partial_t\overline{\zeta}(\alpha,t){\hat{\mathbf{s}}}\right) \quad ; \quad \nu(\alpha,t)=\text{Re}\left(\partial_t\overline{\zeta}(\alpha,t){\hat{\mathbf{n}}}\right).
\end{equation}
The pressure jump across the free sheet is zero, which yields
\begin{equation}\label{eq:pJumpTE}
[p](1,t)=0,
\end{equation}
the boundary condition we use to integrate \eqref{eq:pressure} and obtain
$[p](\alpha,t)$ on the membrane. The derivation of~\eqref{eq:pressure} can be found in~\cite[app.\ A]{mavroyiakoumou2020large}.

The time-dependent boundary conditions are:
\begin{align}
\text{leading edge:}\quad&
    x(-1,t) = -1; \quad y(-1,t) = 0,\label{eq:bc_LE}\\
\text{trailing edge:}\quad&  x(1,t) = x(-1,t)+2\cos \Theta(t); \quad y(1,t) =2\sin \Theta(t).\label{eq:bc_TE}
\end{align}

\begin{figure}[H]
    \centering  
    \includegraphics[width=\linewidth]{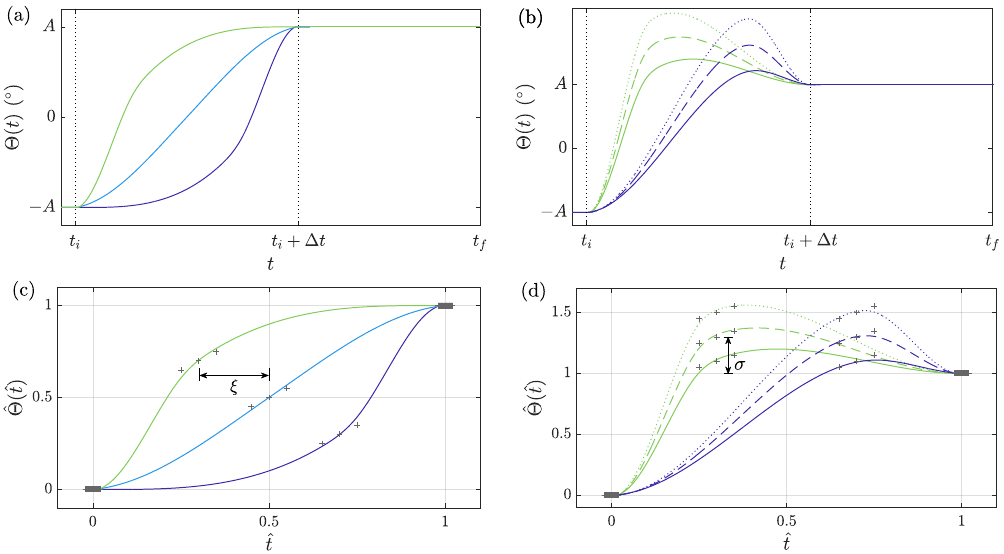} 
    \caption{(a) Three different profiles for angle of attack $\Theta(t)$ as functions of time based on the cubic spline functions $\hat{\Theta}(\hat{t})$ in panel (c). From bottom to top, the three curves correspond to the three offset values $\xi =-0.2$ (dark blue), 0 (blue), and 0.2 (green). (b) Three different profiles for angle of attack $\Theta(t)$ as functions of time with an overshoot parameter that exceeds $A$ by an amount set by $\sigma$ depicted in panel (d) with values equal to $0.1$ (solid line), $0.3$ (dashed line), and $0.5$ (dotted line). The offset values in this case are fixed as $\xi=-0.2$ (dark blue) and $\xi=0.2$ (green).}
    \label{fig:AoAschem}
\end{figure}

Tacking is a strategic maneuver and an objective of this work is to understand how different tacking maneuver strategies lead to different sail membrane dynamics. We focus on the differences in the dynamics between membranes that flip and those that do not flip to the opposite configuration. In figure~\ref{fig:AoAschem} we show the range of tacking kinematics we consider. Panels (a) and (b) depict the angle-of-attack transition kinematics, which are continuous, piecewise-defined functions divided into three distinct parts for the time intervals $[0,t_i]$, $[t_i,t_i+\Delta t]$, and $[t_i+\Delta t,t_f]$, which correspond to times before, during, and after the tacking maneuver, respectively. 

Before tacking $\Theta(t)$ is a constant angle of attack equal to $-A$, during tacking it is equal to a cubic smoothing spline function~\cite{reinsch1967smoothing} computed with the \textsc{Matlab} \texttt{spaps} function~\cite{MATLAB} that goes from $-A$ to $A$ over a time interval that is the total tack time $\Delta t$, and after tacking $\Theta(t)$ is equal to a  constant angle of attack $A$. The details of how $\Theta(t)$ is constructed can be found in appendix~\ref{app:AoAtauAndSigma}.

During the tacking maneuver, a key feature that we control in the angle-of-attack transition kinematics is the concavity of $\Theta(t)$. In figures~\ref{fig:AoAschem}(a) and (c) we consider three cases: a concave-down profile (green), one close to a linear function with zero concavity (light blue), and a concave-up profile (dark blue). The ``offset," $\xi$ in panel (c) to the left or right of $\hat{t}=0.5$, makes the curve concave up or down by controlling the position of the middle gray plusses (explained in detail in appendix \ref{app:AoAtauAndSigma}). In the current work we consider three values of $\xi$: $-0.2$ (dark blue), 0 (blue), and 0.2 (green). 
The effect of~$\xi$ on the membrane dynamics is discussed in detail in \S\ref{sec:offset}. 

Similarly, in figures~\ref{fig:AoAschem}(b) and (d) we show six examples of $\Theta(t)$ profiles by fixing two offset values [$\xi=-0.2$ (dark blue) and $\xi=0.2$ (green)] and using three values of the so-called ``overshoot" parameter $\sigma$: 0.1 (solid lines), 0.3 (dashed lines), and 0.5 (dotted lines). This overshoot parameter $\sigma$ is the vertical distance of the middle gray plusses above 1 and it sets the maximum value of $\Theta(t)$ that the sail attains during the tacking maneuver. We discuss the effect of $\sigma$ on the membrane dynamics in \S\ref{sec:AoA}.
 


\section{Steady membranes at fixed angles of attack}\label{sec:steady}

The dynamics of sail membranes are complicated because the shape of the sail is determined by a pressure distribution which is itself dependent on the profile of the sail~\cite{irvine1979note,kimball2009physics}. This affects also the lift forces on the membranes. We begin by presenting steady membrane shapes for different angles of attack, where the model does not involve tacking (and thus no angle-of-attack transition kinematics or tacking time $\Delta t$) for two values of $R_3$ and a range of $T_0$ values. We use a transient function to slowly change the membrane's angle of attack from zero degrees at $t=0$ to the steady angle of attack $-A$:
\begin{equation}\label{eq:transientVartheta}
    \Theta(t)= -\left(1-e^{-(t/\eta)^3}\right)A,
\end{equation}
where $\eta=5$ controls the ramping-up time and $A$ can admit any value, with the largest value considered here being $20\degre$. 
As in our previous work~\cite{mavroyiakoumou2020large} with 2D fixed-fixed membranes with zero angle of attack, here we find that the 2D sail does not oscillate (i.e. flutter) for these steady nonzero angles of attack. $R_1$ is fixed at $10^{-1}$ but does not affect the steady shapes.

\begin{figure}[H]
    \centering
    \includegraphics[width=.65\textwidth]{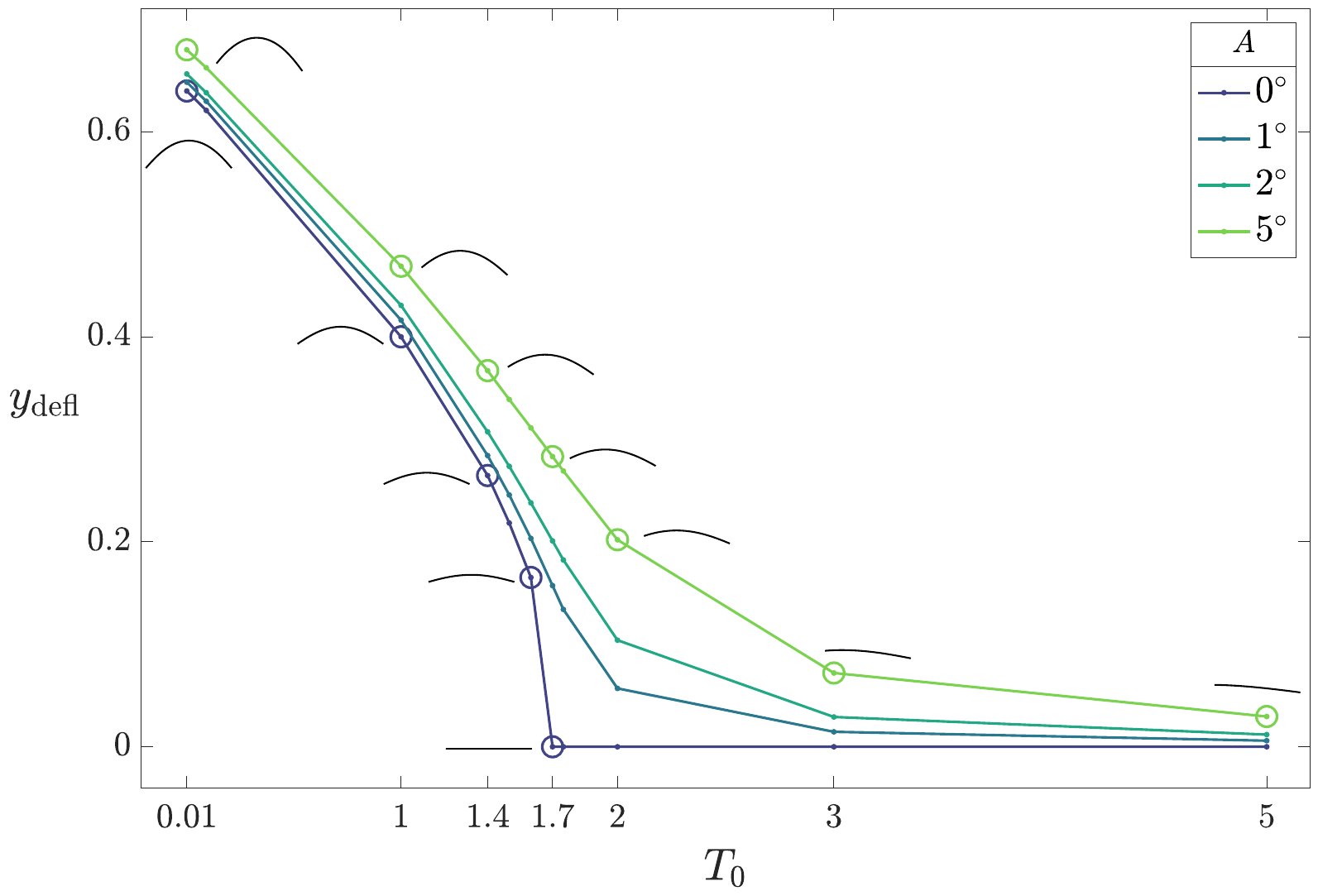}
    \caption{Membrane deflections $y_{\mathrm{defl}}$, defined as the maximum deflection from the line connecting the membrane's endpoints, versus $T_0$ for four values of $A$: $0{\degre}, 1{\degre}, 2{\degre}, 5{\degre}$, ranging from  dark blue to  light green lines, respectively. 
    For each value of $A$ we compute $y_{\mathrm{defl}}$ at 11 values of $T_0$ (0.01, 0.1, 1, 1.4, 1.5, 1.6, 1.7, 1.75, 2, 3, 5) but we only show the membrane snapshots for $A=0\degre$ and $5\degre$ at the $T_0$ values indicated by five blue and seven green circles, respectively. Here $R_1=10^{-1}$ and $R_3 = 10^1$.}
    \label{fig:ydeflvsT0}
\end{figure}

We measure the maximum $y$-deflection from the straight line that connects the membrane's leading and trailing edges and denote it by $y_{\mathrm{defl}}$. We allow the membrane to reach a steady single-hump shape and plot $y_{\mathrm{defl}}$ versus $T_0$ in figure~\ref{fig:ydeflvsT0}. We use four values of $A$: $0{\degre}, 1{\degre}, 2{\degre}, 5{\degre}$; with a dark blue line for $0{\degre}$ through to a light green line for $5{\degre}$. For $A$ equal to $0{\degre}$ and $5{\degre}$ we also show five and seven membrane snapshots each, above and below the corresponding curves at the circled cases. As expected, the membrane at $A=0\degre$ has $y_{\mathrm{defl}}=0$ when $T_0$ is greater than or equal to the threshold value for instability ($T_0\approx 1.7$ in~\cite{mavroyiakoumou2020large,tiomkin2017stability,newman1991stability}). For nonzero values of $A$, the deflection and curvature increase as $A$ increases and $T_0$ decreases.

\begin{figure}[H]
    \centering
    \includegraphics[width=\textwidth]{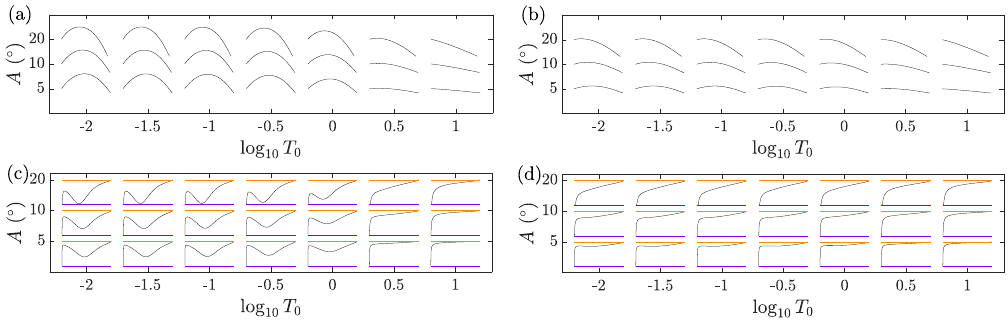}
    \caption{Steady membranes at three values of $A$ ($5{\degre}, 10{\degre}, 20{\degre}$) and seven values of $T_0\in\{ 10^{-2},10^{-1.5},\dots, 10^{1}\}$, with fixed $R_1=10^{-1}$ and (a) $R_3 = 10^1$, (b) $R_3=10^2$. Panels (c) and (d) are the corresponding pressure jump distribution profiles $[p](\alpha,t)$. The orange and purple hash marks denote $[p]=0$ and $-5.4$, respectively.}
    \label{fig:steadyAllICs}
\end{figure}

At zero angle of attack, the steady shape varies with the stretching rigidity $R_3$ as $1/\sqrt{R_3}$ for large $R_3$~\cite{mavroyiakoumou2020large}.
The membrane's steady shape varies with $T_0$ somewhat near the stability boundary, $T_0 \approx 1.7$, then it becomes almost constant as 
$T_0$ decreases below 0.1 (shown in figure~\ref{fig:steadyAllICs}). In figure~\ref{fig:steadyAllICs} we investigate how the steady membrane shapes change with $T_0$ at larger angles of attack than in figure~\ref{fig:ydeflvsT0} and at two different values of stretching rigidity, i.e.\ $R_3$ equal to (a) $10^1$ and (b) $10^2$. In figure~\ref{fig:steadyAllICs}(a) we see that membrane shapes with $T_0\leq 10^{-0.5}$ are almost constant for a fixed angle-of-attack value, because then $R_3$ is dominant in the stretching force. Closer to the stability boundary at $T_0\gtrsim 10^0$ they have a smaller deflection that decreases with increasing $T_0$, because $T_0$ is large enough to be significant in the stretching force. This is still the case in panel (b) with $R_3=10^2$ but is less evident because the deflections are generally smaller. For a fixed pretension, increasing $A$ results in a sail membrane with a higher curvature closer to the leading edge.  

We also plot corresponding pressure jump distribution profiles $[p](\alpha,t)$ in figures~\ref{fig:steadyAllICs}(c) and (d) for $R_3=10^1$ and $10^2$, respectively. The orange and purple hash marks denote $[p](\alpha,t)=0$ and $-5.4$. The Kutta condition makes the pressure jump zero at the trailing edge (according to~\eqref{eq:pJumpTE}). Therefore in all of the $[p](\alpha,t)$ profiles the right end-point is at the orange hash mark. The pressure on a sail's surface is not uniform; near the point of maximum curvature the pressure jump has a local minimum. $[p](\alpha,t)\leq 0$ at all points $\alpha$ since the pressure above the membrane is always smaller than the pressure below. In panel (c) for a fixed $T_0$, the local minimum of the pressure jump becomes more negative as we increase the angle of attack since the membrane becomes more curved and the local minimum point shifts further away from the membrane's midpoint towards the leading edge. In (d) this local minimum in $[p](\alpha,t)$ is usually not present. Instead $[p](\alpha,t)$ is monotonically increasing.



\begin{figure}[H]
    \centering
    \includegraphics[width=\textwidth]{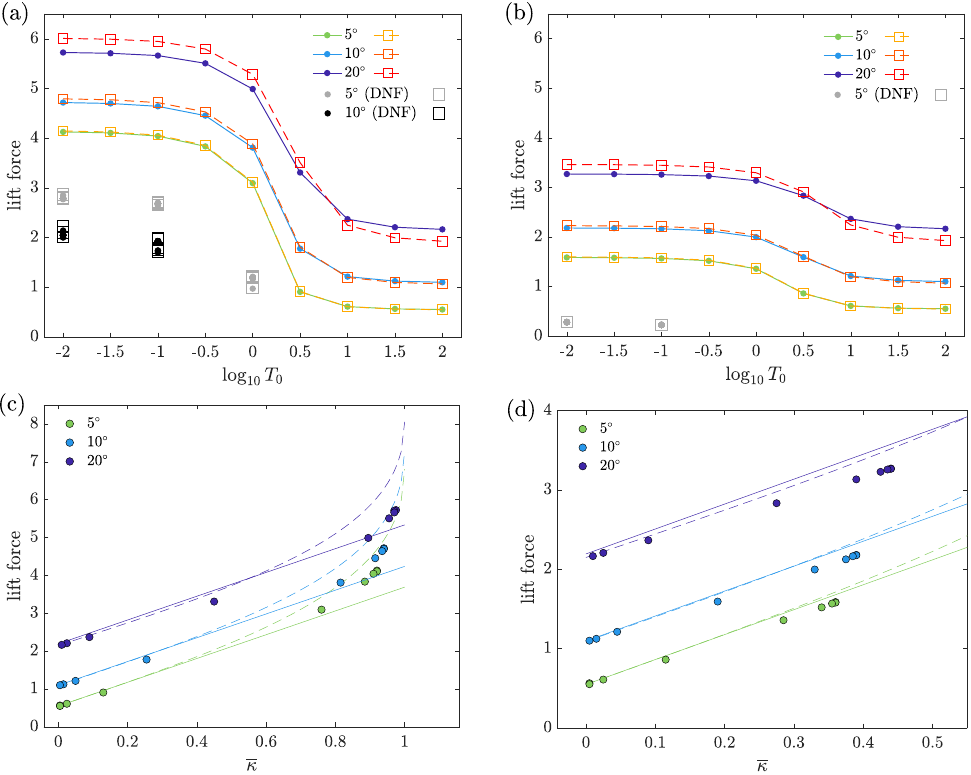}
    \caption{Lift forces for steady membrane shapes and approximate circular arcs. Panels (a) and (b) show lift forces on steady membrane shapes with leading-edge suction component (equation~\eqref{eq:FLES} in appendix~\ref{app:liftAndDrag}) included (green, light blue, dark blue solid lines with dot markers) and without it (yellow, orange, red dashed lines with square markers) versus $T_0$ for steady membranes with three values of $A$: $5{\degre}, 10{\degre}, 20{\degre}$ and (a) $R_3 = 10^1$, (b) $R_3=10^2$. Superposed as circles and squares (in grayscale) are the computed lift forces for all the membranes that after tacking did not flip (DNF) to the opposite curvature. 
    In panels (c) and (d) we show, for each of the three $A$ in (a) and (b), the total lift force versus the corresponding circular arc curvature
    (green, light blue, and dark blue dots) for $R_3$ equal to (c) $10^1$ and (d) $10^2$. The dashed lines show the analytical lift forces (equation~\eqref{eq:liftBatchelor} in appendix \ref{app:liftCircularArc}) for circular arc airfoils (whose solutions exist up to $\overline{\kappa}=1$) and the solid lines show the analytical linearized lift forces~\eqref{eq:linearLiftBatchelor}.}\label{fig:steadyAoAlineLift}
\end{figure}

We plot in figure~\ref{fig:steadyAoAlineLift}(a) and (b) the lift forces, with and without the leading-edge suction force, versus~$T_0$. 
We find that the contribution from the leading-edge suction component is larger at larger angles of attack, shown by the gap between the solid lines with the dot markers and the dashed lines with square markers that increases with angle of attack. The lift forces with $R_3=10^1$ (figure~\ref{fig:steadyAoAlineLift}(a)) are larger than those with $R_3=10^2$ in panel (b) (particularly at $T_0 \ll 1$ where $R_3$ dominates the stretching force) since the steady shapes are more cambered and the pressure jump distributions are larger. 
When we consider the tacking kinematics in~\eqref{eq:bc_LE}--\eqref{eq:bc_TE}, there are two equilibrium states that can be reached depending on the membrane and kinematic parameters. The first is the steady membrane shapes considered so far in this section, reached by the kinematics in~\eqref{eq:transientVartheta}. The second is a set of shapes with smaller mean curvature. These are membranes that did not flip during the tacking kinematics, as shown by examples in the next section (\S\ref{sec:tacking}).

For a given angle of attack, the colored symbols in figure~\ref{fig:steadyAoAlineLift} give lift forces for the equilibrium states with the larger average curvature, while the gray and black symbols (in panels (a) and (b) only) give lift forces for the membranes that did not flip after tacking. We use the acronym DNF to refer to such cases from now on.  In figure~\ref{fig:steadyAoAlineLift}(a) with $R_3=10^1$ there are gray and black symbols because there are membranes that did not flip with both $A=5\degre$ and $10\degre$, whereas in panel (b) there are gray symbols only because only membranes with $A=5\degre$ did not flip. 


In figure~\ref{fig:steadyAoAlineLift}(c) and (d) we compare the computed lift forces for the steady membranes given by colored symbols with the analytical formula (dashed lines) for the lift on a circular-arc membrane that is closest in mean-square deflection to the membrane, versus the constant curvature of the arc, $\bar{\kappa}$. Appendix~\ref{app:liftCircularArc} describes how the circular arc is found for each membrane.

An approximate version of the analytical formula, linearized for small $A$ and small curvature, is given by the solid lines. The formula for the lift forces on the computed membranes is given in appendix~\ref{app:liftAndDrag}. The analytical formula for the circular arc and additional comparisons with membrane shapes and lift values are given in appendix~\ref{app:liftCircularArc}. At $\overline{\kappa} = 1$, the circular arc is a semicircle which deviates significantly from the computed membrane shapes. The tangent angle changes by $\pi$ from the leading to trailing edge which greatly redirects the oncoming flow downward, leading to a large lift force.

In figures~\ref{fig:steadyAoAlineLift}(c)--(d) we show the total lift force (including leading edge suction) versus the curvature of the closest circular arc (by mean-square deflection) for the cases in panels (a) and (b), except for DNF. We superpose as a dashed line with the same color as the corresponding membrane data the lift force given by equation~\eqref{eq:liftBatchelor} (in appendix \ref{app:liftCircularArc}) versus the curvature for the circular arc, defined for 
$\overline{\kappa}\leq 1$. 
The maximum $\overline{\kappa}$ of 1 is attained when the circular arc is a semicircle.
Lastly, we linearize the lift force (equation~\eqref{eq:liftBatchelor}) for small $A$ and small curvature $\overline{\kappa}$, and obtain
\begin{equation}\label{eq:linearLiftBatchelor}
\mathrm{L}_{\ell} = \pi( \overline{\kappa}+2A).    
\end{equation}
We plot~\eqref{eq:linearLiftBatchelor} in panels (c) and (d) as solid lines for each $A$.

The numerically computed drag force (not shown here) is approximately zero, in agreement with D'Alembert's paradox, that states the drag force (including the leading-edge suction part) should be zero for a steady potential flow past a body.






\section{Overview of sail membrane dynamics during tacking}\label{sec:tacking}

We now explain the basic features of membrane dynamics during the tacking simulations. 
Given the steady parameters $R_3$, $T_0$, and $A$ we get steady shapes (a subset of the cases illustrated in figures~\ref{fig:steadyAllICs}(a) and (b)) that we then use as the initial membrane shapes (at $t = 0$) for all the unsteady simulations. These are performed for ranges of values of the dynamical parameters $R_1$, offset $\xi$, and tacking time $\Delta t$ using the kinematics described in figure~\ref{fig:AoAschem} in \S\ref{sec:model}. 

\begin{figure}[H]
    \centering
    \includegraphics[width=\textwidth]{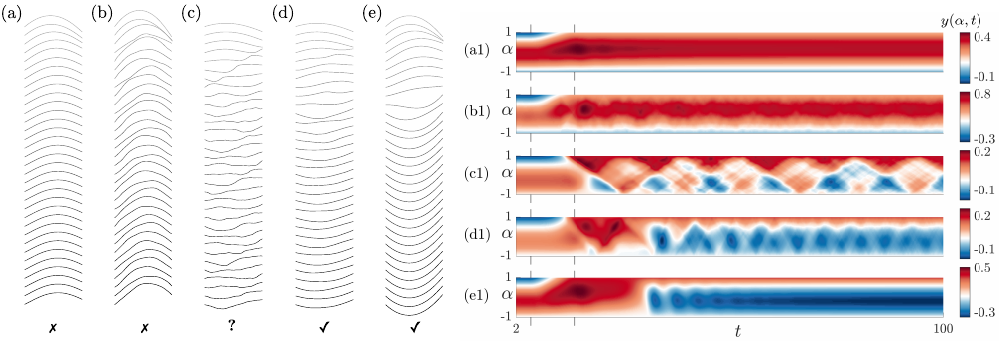}
    \caption{Five examples of typical membrane motions that result from the tacking maneuver. 
    Panels (a)--(e) show sequences of membrane snapshots arrayed vertically, from top to bottom with increasing time, in darkening shades from gray to black at the last times. The color plots in panels (a1)--(e1) show the membrane deflection $y(\alpha,t)$, with $\alpha\in[-1,1]$ and $t$ on the vertical and horizontal axes, respectively, over selected time intervals that illustrate the main dynamics. The tacking maneuver happens when $t\in [5,15]$ and is marked by the black vertical hash marks. 
    The ``\cmark'' symbol indicates flipped membranes, ``\textbf{?}'' indeterminate cases (i.e.\ membranes that oscillate for $t\geq 15$), and ``\xmark'' membranes that did not flip.
    The rest of the membrane and kinematic parameters are $\{R_1,R_3,T_0,A,\xi\}$: (a)--(a1) $\{10^{-1},10^1,10^0,5\degre,0\}$, (b)--(b1) $\{10^{1},10^1,10^{-1},10\degre,0\}$, (c)--(c1) $\{10^{2},10^2,10^{-2},5\degre,-0.2\}$, (d)--(d1) $\{10^{1},10^2,10^{-2},5\degre,-0.2\}$, and (e)--(e1) $\{10^{-2},10^1,10^{0},10\degre,0\}$. The motions in panels (a)--(e) are shown in the supplementary movies named ``Fig7a\dots mp4'' through ``Fig7e\dots mp4,'' respectively.}
    \label{fig:examplesR1}
\end{figure}


In the context of our work a successful tacking maneuver is one that results in a \textit{flipped} membrane.
This corresponds to a sail membrane initially cambered with an angle of $-A$ from the horizontal to its chord (with the entire membrane lying above the chord) that flips to the mirror-image configuration with an angle $A$. That is, the membrane at a time $t$ much greater than the end of the tacking time ($t \gg t_i +\Delta t$) lies entirely below its chord with $y(\alpha,t)\approx-y(\alpha,t_0)$. 
In figure~\ref{fig:examplesR1} we show examples of typical sail membrane motions after tacking, ranging from cases that did not flip ((a) and (b), marked with \xmark) to cases that did flip ((d) and (e), marked with \cmark). 
In panels (a)--(e) we array the membrane snapshots vertically, moving downward with increasing time in darkening shades from gray to black. Here the tacking maneuver happens over a total time $\Delta t = 10$ with the time snapshots shown in the time interval $t\in[2,100]$ which includes the tack $(t\in[5,15])$. For the same time interval we show color plots of the membrane deflection $y(\alpha,t)$ in panels (a1)--(e1), with $-1\leq \alpha\leq 1$ on the vertical axis and $t$ on the horizontal axis. The tack time is marked on each panel by the black vertical hash marks. 
In (a) we show a membrane that eventually after the tack attains a steady single-hump shape. In (b) the membrane never flips to the opposite configuration but during the tacking process undergoes a series of small oscillations about the single-hump shape that decay away at long times. In (c) we observe an unusual state where a traveling wave is initiated on the membrane near the end of the tacking time interval and persists without decaying for the remaining simulation time (up to $t=500$ but in figure~\ref{fig:examplesR1} shown up until $t=100$). Since this state does not converge to either the flipped or non-flipped states, it is marked \textbf{?}. In panel (d) the membrane flips but its motion is characterized by small oscillations about the flipped steady shape. The membrane converges to a steady shape at later times. 
Finally, in panel (e) the membrane flips to the opposite configuration and becomes steady.

\begin{figure}[H]
    \centering
     \includegraphics[width=\textwidth]{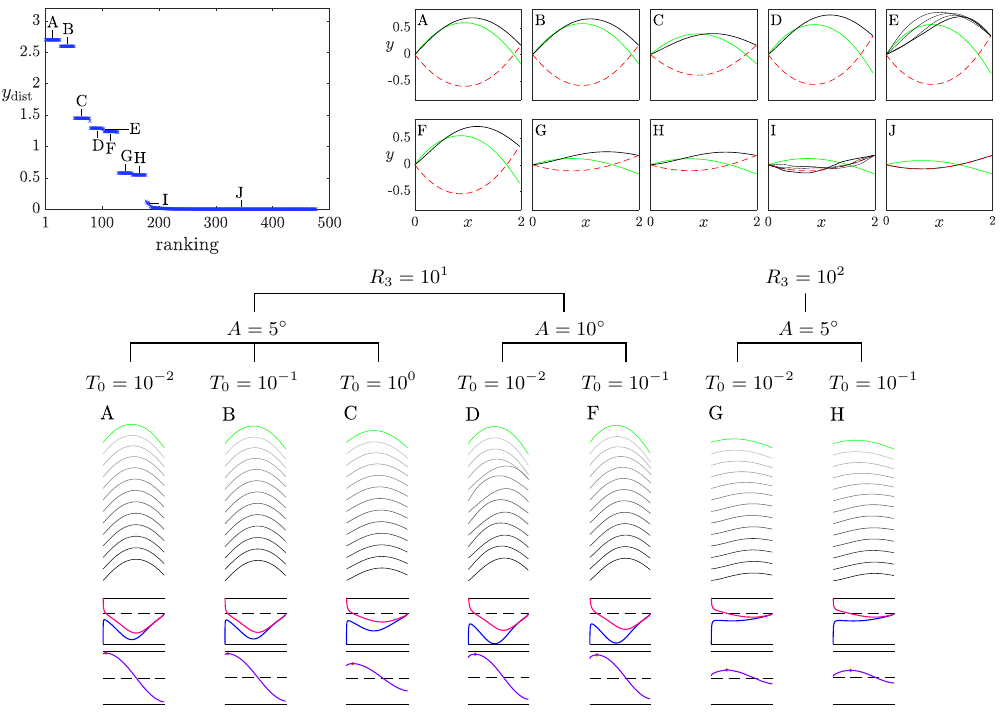}
    \caption{Top left panel: Ranking of all cases in the six-parameter space based on $y_{\mathrm{dist}}$, defined in~\eqref{eq:ydist}. Membranes with the largest $y_{\mathrm{dist}}$ (A) are furthest from flipping whereas membranes with $y_{\mathrm{dist}}\approx 0$ have flipped (J). We use letters A through J to denote a representative example from each group of $y_{\mathrm{dist}}$ values. Top right panels A--J: Membrane shapes at $t=t_i$ when $\Theta=-A$ (green line) and at $t=t_f$ when $\Theta=A$ (black line) for A--J. The red dashed line shows the mirror image of the initial configuration which is the shape the membrane would have if it flipped in all cases. E and I have $R_1=10^2$ and still oscillate at $t = t_f$, so we show four membrane snapshots at different times to demonstrate the range of motions. Bottom portion: Membrane snapshots arrayed vertically moving downward with increasing time for A--C ($R_3=10^1$ and $A=5\degre$), D--F ($R_3=10^1$ and $A=10\degre$), and G--H ($R_3=10^2$ and $A=5\degre$). Below each sequence of snapshots we also show the pressure jump distribution profile at $t=t_i$ and $t_f$, as a blue and magenta line respectively. The black solid hash marks denote $[p]=2$ and $-4$, and the dashed ones $[p]=0$. At the very bottom, we plot the local tangent angle $\theta(\alpha,t_f)$ as a purple line, together with its extremum point as an orange dot. The black solid hash marks show $-50\degre$ and $50\degre$ and the dashed hash marks show $\theta =0$.}
    \label{fig:noflipping}
\end{figure}

Having presented examples of membrane dynamics that range from not flipping to flipping in figure~\ref{fig:examplesR1}, we now quantify this range and classify all the membranes according to their location along it. 
We define a metric $y_{\mathrm{dist}}$, to describe the distance of the time- and space-averaged membrane profile from $-y(\alpha,t_0)$, the flipped state, which is the mirror image of the initial condition. Its definition is:
\begin{equation}\label{eq:ydist}
 y_{\mathrm{dist}}:= \displaystyle \frac{1}{t_f/2}\int_{t_f/2}^{t_f}\left(\frac{1}{2}\int_{-1}^1 \left|y(\alpha,t)+y(\alpha,t_0)\right|\partial_{\alpha} s \,\d\alpha\right)\d t \bigg/
|y(1,t_0)|,
\end{equation}
where $t\in[t_{f}/2,t_f]$ is taken over the last several hundred time units, $t_0$ is the initial time in the simulation, and $|y(1,t_0)|$ is a normalization factor. In ranking the membranes according to $y_{\mathrm{dist}}$ we find that membranes that remain in the non-flipped states after tacking assume a few kinds of motions, and in figure~\ref{fig:noflipping} we display one of each kind, in descending order of $y_{\mathrm{dist}}$~\eqref{eq:ydist}. We also analyze the membrane shape profiles, the pressure distributions, and $\theta(\alpha,t)$ before and after tacking.


The top left panel of figure~\ref{fig:noflipping} shows the ranking of the membranes in descending order of $y_{\mathrm{dist}}$, for all the computed cases in the six parameter space of $\{R_1,R_3,T_0,A,\Delta t,\xi\}$. We consider 5 values of $R_1$: $[10^{-2},10^{-1},10^{0},10^1,10^2]$, 2 values of $R_3$: $[10^{1}, 10^{2}]$, 3 values of $T_0$: $[10^{-2},10^{-1},10^0]$,  3 values of $A$: $[5\degre,10\degre,20\degre]$, 2 values of $\Delta t$: $[10,80]$, and 3 values of $\xi$: $[-0.2,0,0.2]$, giving 540 cases in total. 64 of these cases failed to converge to the desired error tolerance at a time $t < 150$ and were unable to proceed to larger $t$. This mostly occurred at large $R_1$ ($10^2$) with small $\Delta t$ (10) and small $\xi$ ($-0.2$ and 0). In figure~\ref{fig:noflipping} we omit the 64 cases that did not run for more than 150 time units because 150 is approximately two times the maximum $\Delta t$ considered here, so one would not expect steady state dynamics much before this time. We analyze only the remaining 476 cases, and all of these ran for $t \geq 440$, at least five times $\Delta t$, long enough to reach the steady-state dynamics.
For these 476 cases $y_{\mathrm{dist}}$ values are plotted in the top left panel. The larger the value of $y_{\mathrm{dist}}$, the further from flipping the membrane is. The extreme case of $y_{\mathrm{dist}} \approx 0$ corresponds to flipped membranes (since then $y(\alpha,t)\approx -y(\alpha,t_0)$ for $t \in [t_f/2,t_f]$ in~\eqref{eq:ydist}). We use letters A--J to denote a representative example from each of the ten groups of $y_{\mathrm{dist}}$ values. We also use this letter to refer to the group as a whole. The vertical and horizontal hash marks connect each letter to the value of the example considered.

The top right $2\times 5$ panels show: the membrane shapes at the initial time before tacking (green line) superposed with the membrane shape at the final time after tacking (black line), and the membrane shape if it were to flip (red dashed line). The latter is the same as the mirror image of the green line.  
Cases E and I have a large membrane mass density ($R_1=10^2$) and still oscillate at the end of the simulation, so we show three additional time snapshots of the membrane shapes to illustrate the range of motions. Examples I and J correspond to membranes that flipped. 

At the bottom portion of figure~\ref{fig:noflipping} we show sequences of membrane snapshots arrayed vertically, from top to bottom with increasing time, in darkening shades from gray to black at the later times for cases A--D and F--H (i.e.\ membranes that did not flip).
The motions (and values of $y_{\mathrm{dist}}$) of A and B are similar; these highly curved membranes maintain an almost unchanged profile during the tacking maneuver (i.e.\ while their angle of attack increases from $-A$ to $A$). For cases A and B the steady parameters $R_3$, $T_0$, and $A$ are similar: they both have a small stretching rigidity 
($R_3=10^1$) and a small $A$ ($5\degre$), while the pretension $T_0$ is $10^{-2}$ and $10^{-1}$, respectively.
Case C has the same $R_3$ and $A$ as cases A and B but a larger $T_0$ ($10^0$). This results in a membrane shape with a smaller mean curvature and a significantly smaller value of $y_{\mathrm{dist}}$, making it closer to the flipped state.
The membrane deflection in D and F is about the same as for A and B but the trailing edge deflection is about twice as much as for A--B, so $y_{\mathrm{dist}}$ for D--F is about half as much as for A--B. However, for G--H the trailing edge deflection is about the same as in A--B but the membrane deflection is much smaller which results in a much smaller $y_{\mathrm{dist}}$.

Below each sequence of membrane snapshots we show the pressure jump distributions at $t=t_i$ (blue lines) and at $t_f$ (magenta lines). The black solid hash marks are used to denote $[p]=2$ and $-4$, and the dashed ones to denote $[p]=0$. The distributions for G and H are more uniform along $\alpha$ in comparison to A--F where the difference between the pressure above and below the membrane (i.e.\ $[p]=p_+-p_-$) is very large near membrane's maximum curvature point. 
At the very bottom of figure~\ref{fig:noflipping} (below the $[p](\alpha,t_i)$ and $[p](\alpha,t_f)$ profiles) we plot the local tangent angle $\theta(\alpha,t_f)$ as a purple line, together with the extremum point of $\theta(\alpha,t_f)$ as an orange dot, an inflection point on the membrane. Going from left to right at the very bottom of figure~\ref{fig:noflipping} we observe that this orange dot moves from the membrane's leading edge towards its midpoint $\alpha=0$. 
The closer this point is to the membrane's midpoint, the closer the membrane is to flipping (cases G and H). 


This ranking method reveals that the ten groups of $y_{\mathrm{dist}}$ values correspond to different groups of parameters that determine the steady shapes of the membranes, i.e. $R_3$, $T_0$, and $A$. Within each of these groups there is a variety of the remaining three parameters: $R_1$, $\Delta t$, and $\xi$; parameters that mainly control the membranes' dynamics. Membranes with $R_3=10^1$, $A=5\degre$ and $T_0=10^{-2}$, $10^{-1}$, and $10^0$ are found in groups A--C, respectively. Instead membranes in groups D and F have the same value of stretching rigidity $R_3$ $(10^1)$ as groups A--C but a larger $A$ ($10\degre$), and the pretensions are $T_0=10^{-2}$ and $10^{-1}$. Groups G and H have large $R_3$ ($10^2$), small $A$ ($5\degre$) and $T_0=10^{-2}$ and $10^{-1}$. Both groups E and I have $R_1=10^{2}$ and are still oscillating at $t = t_f$. Group I has $R_1=10^2$ and a wide variety of the other parameters. Finally, group J (i.e., flipped membranes) consists of all the membranes with $A=20\degre$ as well as membranes with other combinations of other parameters including some with $A=5$ and $10\degre$.

\begin{table}[H]
\centering
\begin{subfloat}[$\mathrm{mean}(y_{\mathrm{dist}})$ with $R_3=10^1$.]{
\renewcommand{\arraystretch}{1} 
\setlength{\tabcolsep}{0.2cm} 
\begin{tabular}{c||c|c|c}
\backslashbox{$T_0$}{$A$}& $5\degre$ &  $10\degre$ &  $20\degre$ \\ \hline\hline
& & &\\[-.3cm]
$10^{0}$& 1.36 (C) & $4.88\times 10^{-3}$ & $2.60\times 10^{-3}$\\[.05cm]
$10^{-1}$& 2.60 (B) & 1.19 (F) & $2.80\times 10^{-3}$\\[.05cm]
$10^{-2}$& 2.70 (A) & 1.29 (D) & $3.05\times 10^{-3}$\\
\end{tabular}}
\end{subfloat}
\hspace{1.7cm}
\begin{subfloat}[Percentage of cases that belong in groups~I and J with $R_3=10^1$.]{
\setlength{\tabcolsep}{0.3cm} 
\begin{tabular}{c||c|c|c}
\backslashbox{$T_0$}{$A$}& $5\degre$ &  $10\degre$ &  $20\degre$ \\ \hline\hline
& & &\\[-.3cm]
$10^{0}$& $6.67\%$ & $100\%$ & $100\%$\\[.05cm]
$10^{-1}$& $0\%$ & $3.85\%$ & $100\%$\\[.05cm]
$10^{-2}$& $0\%$ & $0\%$ & $100\%$\\
\end{tabular}}
\vspace{.2cm}
\end{subfloat}\\
\begin{subfloat}[$\mathrm{mean}(y_{\mathrm{dist}})$ with $R_3=10^2$.]{
\renewcommand{\arraystretch}{1} 
\setlength{\tabcolsep}{0.2cm} 
\begin{tabular}{c||c|c|c}
\backslashbox{$T_0$}{$A$}& $5\degre$ &  $10\degre$ &  $20\degre$ \\ \hline\hline
& & &\\[-.3cm]
$10^{0}$& $9.03\times 10^{-3}$ & $3.93\times 10^{-3}$ & $1.45\times 10^{-3}$\\[.05cm]
$10^{-1}$& $4.72\times 10^{-1}$ (H) & $2.74\times 10^{-3}$ & $1.93 \times 10^{-3}$\\[.05cm]
$10^{-2}$& $5.04\times 10^{-1}$ (G) & $4.16\times 10^{-3}$ & $1.98\times 10^{-3}$\\
\end{tabular}}
\end{subfloat}
\hspace{1cm}
\begin{subfloat}[Percentage of cases that belong in groups I and J with $R_3=10^2$.]{
\renewcommand{\arraystretch}{1} 
\setlength{\tabcolsep}{0.3cm} 
\begin{tabular}{c||c|c|c}
\backslashbox{$T_0$}{$A$}& $5\degre$ &  $10\degre$ &  $20\degre$ \\ \hline\hline
& & &\\[-.3cm]
$10^{0}$& $100\%$ & $100\%$ & $100\%$\\[.05cm]
$10^{-1}$& $14.29\%$ & $100\%$ & $100\%$\\[.05cm]
$10^{-2}$& $13.79\%$ & $100\%$ & $100\%$\\
\end{tabular}}
\vspace{.2cm}
\end{subfloat}
\caption{Tables showing in $T_0$-$A$ space for two values of $R_3$:
(a,b) $10^1$ and (c,d) $10^2$, values of $\mathrm{mean}(y_{\mathrm{dist}})$ (in (a) and (c)) and the percentage of cases with a specific $(R_3,T_0,A)$ combination that flipped, i.e. that belong to groups I and J (in (b) and (d)). The computed values are for all combinations of $R_1$ values and the other kinematic parameters, $\xi$ and $\Delta t$, that computed successfully. When $\mathrm{mean}(y_{\mathrm{dist}})=\mathcal{O}(10^{-3})$, the corresponding percentage of cases that flipped is 100\%. If instead  $\mathrm{mean}(y_{\mathrm{dist}})= \mathcal{O}(10^{-1})$ or $\mathcal{O}(10^{0})$, most or all membranes did not flip. For these combinations of $(R_3,T_0,A)$ we also indicate the letter of the group they belong to, from the bottom of figure~\ref{fig:noflipping}.  } \label{tab:MeanAndStdFlipValsR310and100}
\end{table}

In figure \ref{fig:noflipping} we showed that the computed membrane motions could be sorted into 10 groups, and groups of membranes that did not flip were associated with particular values of $(R_3,T_0,A)$. 
However, at some $(R_3,T_0,A)$ there were mixtures of cases that did and did not flip. In table \ref{tab:MeanAndStdFlipValsR310and100} we show more clearly the relation between $y_\mathrm{dist}$ and $(R_3,T_0,A)$ and at each value the percentage of cases that did not flip. Specifically, 
we compute the mean of $y_\mathrm{dist}$ in $T_0$-$A$ space, presented in different sub-tables: table \ref{tab:MeanAndStdFlipValsR310and100}(a) and (c) for the two values of $R_3$ ($10^1$ and $10^2$, respectively).
This measure, together with the percentage of cases for each combination of $R_3$, $T_0$, and $A$ that flipped (i.e. belong in groups I and J) (in sub-tables \ref{tab:MeanAndStdFlipValsR310and100}(b) and (d)), reveal where in $R_3$-$T_0$-$A$ space there are more cases of membranes that flipped given all combinations of the three remaining parameters---$R_1$, $\Delta t$, and $\xi$---that control the dynamics leading to the final state.

There are two groups of cases that emerge. The first group has very small $\mathrm{mean}(y_{\mathrm{dist}})$ and 100\% of the membranes flip. This occurs at $A=20\degre$ with any $R_3$ and $T_0$, at $A=10\degre$ with large $R_3$ or large $T_0$, and at $A=5\degre$ with both $R_3$ and $T_0$ large. The second group, consisting of the remaining cases, has $\mathrm{mean}(y_{\mathrm{dist}})\approx 0.4$--2.7 and only 0--15\% of the membranes flipping.
Membranes in this second group have profiles similar to A--H, shown in figure~\ref{fig:noflipping}. 
Dynamical/kinematic parameters ($R_1$, $\Delta t$, and $\xi$) can affect whether flipping occurs at these $(R_3,T_0,A)$ values. 
In general, however, flipping is largely determined by $(R_3,T_0,A)$. Table~\ref{tab:MeanAndStdFlipValsR310and100} shows that, without exception, flipping is more likely to occur at large $T_0$, large $R_3$, and large $A$.

We have presented some of the basic aspects of sail membrane dynamics during tacking. 
In the next two sections we study in detail the effects of the seven main parameters on the dynamics. The parameters can be divided into two main categories: those that determine the properties of the sail membrane and those that control the kinematics of the tacking process. In the former we have the membrane mass density $R_1$, the membrane stretching rigidity $R_3$, and the membrane pretension~$T_0$. In the latter we have the final angle of attack $A$, the total time of tacking $\Delta t$, and the angle-of-attack transition kinematics, determined by $\xi$ and $\sigma$ (as in figure~\ref{fig:AoAschem}).

\section{Results 1: effects of membrane parameters}\label{sec:membraneParameters}

In this section we study separately the effects of the membrane mass density $R_1$, the stretching rigidity $R_3$, and the pretension~$T_0$. These parameters determine the properties of the sail membrane rather than the kinematics of the tacking maneuver. 
The sail motions are strongly dependent on the membrane properties. 
Previous sail studies~\cite{le1999unsteady,newman1991stability} considered $R_1\in[0,6]$, $R_3\in[10,1000]$, and $T_0\in[0,2]$; here we consider values within and somewhat beyond these ranges.

\subsection{Effect of membrane mass density}
\begin{figure}[H]
\centering
    \includegraphics[width=\textwidth]{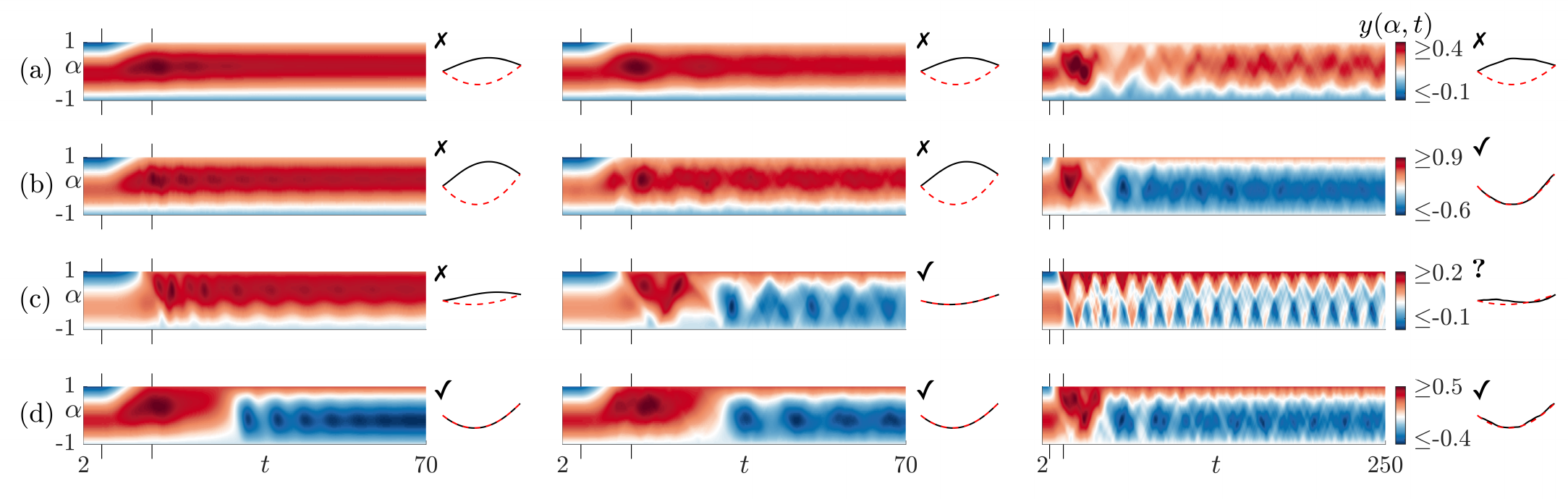}
    \caption{(Effect of $R_1$.) Color plots showing the membrane deflection $y(\alpha,t)$ with $\alpha$ and $t$ on the vertical and horizontal axes respectively, over time intervals that cover the motions before, during, and after the tacking maneuver: $t\in [2,70]$ in columns 1--2 and $t\in[2,250]$ in column 3. 
    The time interval over which the tacking maneuver is performed ($\Delta t=10$ in all cases) is marked by the two black vertical hash marks. The ``\cmark'' symbol indicates flipped membranes, ``\textbf{?}'' indeterminate cases (i.e.\ membranes that oscillate for $t\geq t_i+\Delta t$), and ``\xmark'' membranes that did not flip. Next to each color plot we show the membrane snapshot at $t_f$ as a black curve and $-y(\alpha,t_0)$ as a red dashed line.
    Each column (within each panel (a)--(d)) has a different value of $R_1$: $10^0,10^1,10^2$, from left to right. The remaining four parameters $\{R_3,T_0,A,\xi\}$ are: (a) $\{10^1,10^0,5\degre,0\}$, (b) $\{10^1,10^{-1},10\degre,0\}$, (c) $\{10^2,10^{-2},5\degre,-0.2\}$, and (d) $\{10^1,10^{0},10\degre,0\}$. The motions in each of the color plots in panels (c) and (d) are shown in the supplementary movies named ``Fig9c1\dots mp4'' through ``Fig9d3\dots mp4,'' respectively.}
    \label{fig:effectOfR1}
\end{figure}

In figure~\ref{fig:effectOfR1} we characterize the typical membrane dynamics when we increase the membrane mass density from $10^{0}$ to $10^2$ (from left to right) while keeping the remaining five parameters ($R_3,T_0,A,\Delta t,\xi$) fixed. The behavior at $R_1 < 10^{0}$ is approximately the same as with $R_1=10^0$ (the first column) because the membrane mass is too small to affect the dynamics. In each of the panels (a)--(d), we show color plots of the membrane deflection $y(\alpha,t)$ as a function of the material coordinate $\alpha$ (on the vertical axis) and time $t$ (on the horizontal axis). In panels (a)--(d) the total tacking time is fixed as $\Delta t=10$ but the combination of the remaining four parameters is different in each case. The parameters are chosen to reflect different sequences of motions that are observed as $R_1$ increases; ranging from a case in which membranes do not flip for any $R_1$ (panel (a)) to a case in which they flip for all $R_1$ (panel (d)). The cases that did not flip belong to group C (in panel (a)), group~F (the first two in panel (b)), and group G (the first in panel (c)), and the rest belong to groups I and J. 

In figure~\ref{fig:effectOfR1}(a) the stretching rigidity is $R_3=10^1$, the pretension is $T_0=10^0$, the final angle of attack is $A=5\degre$, and the offset is $\xi=0$. With these membrane and kinematic parameters, the membrane does not flip after tacking for any of the $R_1$ values.  At small $R_1$ the membrane undergoes small oscillations about a single-hump shape and by $t\approx 50$ (several time units after the tacking maneuver ends) it tends to a steady shape with a single hump and with the maximum deflection point closer to the membrane's trailing edge. As $R_1$ is increased to $10^1$ (column 2) slightly larger oscillations occur that correspond to a sharpening and broadening of the maximum deflection point. At the largest $R_1$ ($10^2$) the membrane almost flips but instead remains in the non-flipped state and exhibits significantly larger oscillations that persist at late times. In panel (b) $R_3$ and $\xi$ are the same as in (a) but $T_0$ is smaller ($10^{-1}$) and $A$ is larger ($10\degre$). With these parameters, the membrane motions at small $R_1$ are similar to those in (a) but with larger oscillations about the single-hump shape, and larger overall deflections. When $R_1$ is increased to $10^2$ (column 3) the membrane flips.

In panel~(c) we consider a more rigid membrane whose $R_3$ is set to $10^2$. The other parameters are $T_0=10^{-2}$, $A=5\degre$ and $\xi=-0.2$. This smaller value of $\xi$ (as explained in \S\ref{sec:model}) gives a $\Theta(t)$ that is concave up during the tacking maneuver. Now, membranes with $R_1\leq 10^0$ still do not flip (as in (a) and (b)) and have small oscillations but the membranes converge to a steady shape at later times. Increasing the value of $R_1$ to $10^1$ (column 2) leads to a flipped membrane which still oscillates several time units after the tacking maneuver is finished. With $R_1=10^2$ we have the (same) unusual state shown in figure~\ref{fig:examplesR1}(c1) where the membrane is relatively steady during tacking but then starts to oscillate with a traveling wave motion that persists for the remaining simulation time. In panel (d) the parameters are the same as in (a) except $A$ is $10\degre$ instead of $5\degre$. In this case, the membrane flips for any of the $R_1$ values. The oscillation frequencies generally decrease with increasing membrane mass density. The decreased frequency corresponds to a wider horizontal spacing between the repeated features in the color plots. We found similar behaviors and dependencies of oscillation frequencies on membrane mass densities in our previous works~\cite{mavroyiakoumou2020large,mavroyiakoumou2021dynamics,mavroyiakoumou2022membrane,mavroyiakoumou2023spanwise}.

The examples in figure~\ref{fig:effectOfR1} indicate the range of effects of $R_1$ on flipping. In general, $R_1$ only affects flipping in a small number of cases, and these occur in the boxes of table I(b) and (d) that are slightly above 0\%. In these cases, flipping does not occur for small $R_1$ ($\leq 10^0$) but may occur for $R_1 \geq 10^1$ or $10^2$. These cases may occur with all $\xi$ values but mostly have $\Delta t = 10$, not 80. With large $R_1$ membrane inertia seems to allow the membrane to maintain the momentum from the tacking motion long enough to reach the flipped state. At the largest $R_1$ of $10^2$ a few cases still had sustained oscillations at $t = 500$ and flipping was somewhat indeterminate.






\subsection{Effect of membrane stretching rigidity and pretension}

\begin{figure}[H]
    \centering
    \includegraphics[width=\textwidth]{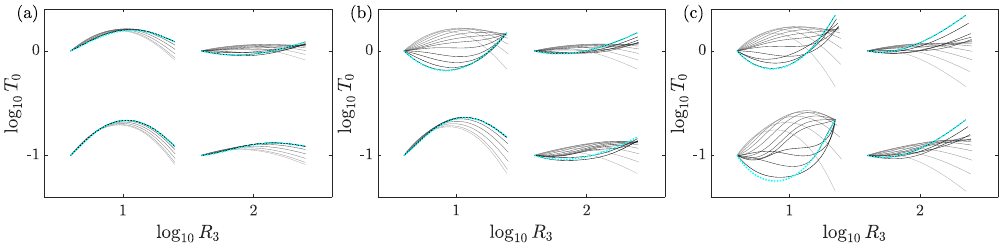}
    \caption{(Effect of $R_3$ and $T_0$.) Sequence of membrane snapshots ranging from gray at early times to black at late times in $R_3$-$T_0$ space for $A$  equal to (a) $5\degre$, (b) $10\degre$, and (c) $20\degre$. The dotted cyan line in each case denotes the membrane snapshot at the final simulation time $t_f=500$. Here $R_1=10^{-2}$, $\Delta t=80$, and $\xi=0$.}
    \label{fig:effectOfR3andT0}
\end{figure}

In table~\ref{tab:MeanAndStdFlipValsR310and100} we showed that there is a greater tendency for sail membranes to flip with larger stretching rigidity, pretension, and angle of attack.
We seek to understand their individual effects on the sail dynamics by studying the membrane motions for different combinations of small and large values of $R_3$ and $T_0$ (for fixed~$A$).
In figure~\ref{fig:effectOfR3andT0} we show sequences of membrane snapshots in $R_3$-$T_0$ space, across three values of $A$: (a) $5\degre$, (b) $10\degre$, and (c) $20\degre$. The other parameters are $R_1=10^{-2}$, $\Delta t=80$, and $\xi=0$. Shades of gray increase from light to dark as the time increases from $t_i$ to $t_i+1.5\Delta t$ (with times chosen to illustrate the full range of motions), while the dotted cyan lines show the membrane shapes at the final simulation time $t_f=500$.

At the smallest $A$ considered here ($5\degre$, in panel (a)) the only membrane that flips is the one with both $R_3$ and $T_0$ large, $10^2$ and $10^0$, respectively. 
For this small value of angle of attack, if either $T_0$ or $R_3$ (or both) are not sufficiently large, then the membrane remains a single-hump shape throughout the simulation, including the time interval over which the tacking takes place. 
When $(R_3,T_0)=(10^1,10^{-1})$ the membrane shape is almost fore-aft symmetric and steady at late times. When either the stretching rigidity or the pretension are large, i.e. when $(R_3,T_0)$ is equal to $(10^2,10^{-1})$ or $(10^1,10^0)$, the membrane snapshots show significant deformations closer to the leading edge where the concavity changes sign during tacking and an inflection point occurs. 


Increasing the angle of attack to $10\degre$ in figure~\ref{fig:effectOfR3andT0}(b) yields membranes that now flip when {\it either} $R_3$ {\it or} $T_0$ is large. When both of these membrane parameters are small ($R_3=10^1$ and $T_0=10^{-1}$) the membrane remains in the non-flipped state with a single-hump profile that is steady at late times and the membrane has an interior inflection point, similar to some cases in (a). The three cases shown of membranes that flip have considerably different motions during the tacking maneuver. 

In panel (c) with $A=20\degre$ all membranes flip. With a stretching rigidity $R_3$ equal to $10^1$ the membrane can deform more than when it is equal to $10^2$ and so the maximum deflection point reached by the membrane during the tacking maneuver is significantly larger in the left column of panel (c). A comparison between $T_0=10^{-1}$ and $10^0$ indicates that $T_0$ also affects the maximum deflection point in the same way.
Therefore, for a fixed $A$, the sequence of membrane snapshots reveals that membranes that are less curved are more likely to flip.

We also compute the motions in figure~\ref{fig:effectOfR3andT0} across ranges of $R_1$ ($10^{-2}$, $10^1$, and $10^2$), $\Delta t$ (10 and 80), and $\xi$ ($-0.2$, 0, and 0.2) and find that flipping occurs at the same combinations of $R_3$ and $T_0$ as in figure~\ref{fig:effectOfR3andT0}, except in a few cases with $R_1 = 10^1$ and $10^2$, with $T_0 = 10^{-1}$ and $R_3 = 10^2$, corresponding to the lower right corner of figure~\ref{fig:effectOfR3andT0}(a). The nonflipping case there changes to flipping with large membrane mass. To summarize, large $R_3$ and $T_0$ both lead to smaller membrane deflections and result in flipped membranes. When these two parameters are small, then $A$ must be large for the membranes to flip.



\section{Results 2: effects of kinematic parameters}\label{sec:kinematicParameters}
We now focus on the parameters that control $\Theta(t)$, the kinematics of the tacking maneuver. These are: $A$, $\Delta t$, $\xi$, and~$\sigma$. For the membranes that flip we consider how quickly they do so using three different flip-time measures; all of them based on the following ratio:
\begin{equation}\label{eq:flipTimeRatio}
   \Upsilon(t) ={\displaystyle\int_{-1}^1 (y(\alpha,t_f)-y(\alpha, t))^2 \d \alpha} \bigg/ { \displaystyle\int_{-1}^1 (2y(\alpha,t_f))^2 \d \alpha},
\end{equation}
which is the relative square difference between the membrane shape at time $t$ and at the final time $t_f$ (the flipped shape). The first flip-time measure  $t_1^{\mathrm{flip}}:=\int_0^{t_{f}}\Upsilon(t)\,\d t$ gives the time-integrated deviation of the membrane shape from the final flipped shape. Therefore, the sooner $y(\alpha,t)$ approaches $y(\alpha,t_f)$ the smaller $t_1^{\mathrm{flip}}$ will be.
The other two flip-time measures are defined as the latest times at which the ratio $\Upsilon(t)$ exceeds 0.01 and 0.001, respectively. These times are denoted by $t_2^{\mathrm{flip}}$ and $t_3^{\mathrm{flip}}$. In some cases the membrane approaches the flipped shape quickly but does not become very close to it, so $t_1^{\mathrm{flip}}$ is relatively small but $t_2^{\mathrm{flip}}$ or $t_3^{\mathrm{flip}}$ may be large, and it is useful to consider all three.

\subsection{Effect of final angle of attack}\label{sec:AoA}

\begin{figure}[H]
\centering
\includegraphics[width=\textwidth]{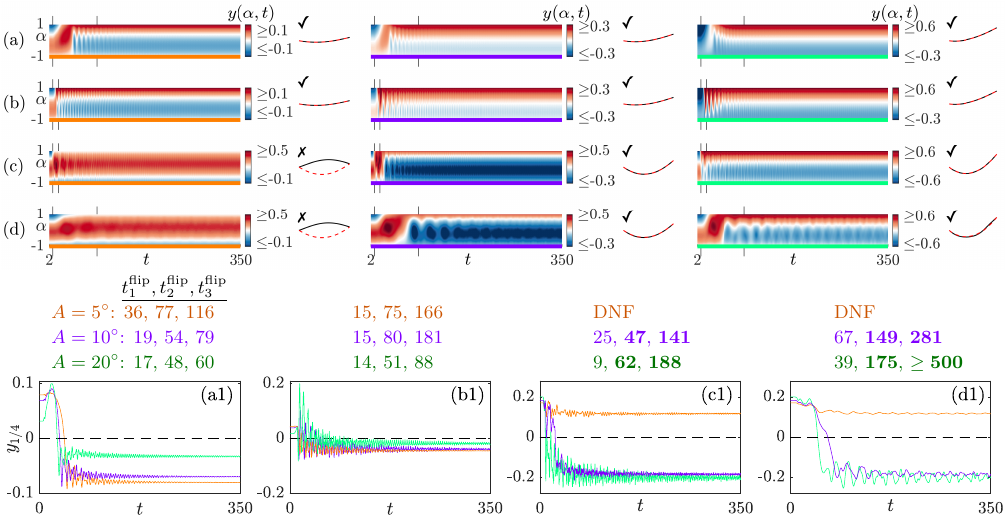}

    \caption{(Effect of $A$.) The color plots show $y(\alpha,t)$.
    Each column corresponds to a different value of $A$: $5\degre$ (left),  $10\degre$ (middle), and \ $20\degre$ (right); distinguished by the orange, purple, and green horizontal time-axes. The rest of the parameters $\{R_1,R_3,T_0,\Delta t,\xi\}$ are: (a) $\{10^1,10^2,10^0,80,0.2\}$, (b) $\{10^1,10^2,10^0,10,-0.2\}$, (c) $\{10^1,10^1,10^0,10,0.2\}$, and (d) $\{10^2,10^1,10^0,80,0.2\}$. Below the color plots, we give the computed flip times $t_1^{\mathrm{flip}}$, $t_2^{\mathrm{flip}}$, and $t_3^{\mathrm{flip}}$ as lists of three numbers, colored according to the $A$ value they correspond to. At the bottom, in panels (a1)--(d1) we plot $y_{1/4}(t)$, with the corresponding color for each $A$ and with the same time range as for the color plots. The black dashed lines show the zero line.
    }
    \label{fig:AoAEffect}
\end{figure}

We begin by considering the angle of attack $A$; the angle that the sail's chord makes with the horizontal at $t=t_f$. 
In figure~\ref{fig:AoAEffect} we focus on three values of $A$: $5\degre$ (left column), $10\degre$ (middle column), and $20\degre$ (right column). In rows (a)--(d), for each of these $A$ we show typical membrane motions as color plots of membrane deflection $y(\alpha,t)$ for $t\in[0,350]$. We color the horizontal time-axes in (a)--(d) to distinguish them more easily:  orange for $A=5\degre$, purple for $10\degre$, and green for $20\degre$.
Below the color plots, we give the computed flip times in terms of the three measures $t_1^{\mathrm{flip}}$, $t_2^{\mathrm{flip}}$, and $t_3^{\mathrm{flip}}$ as lists of three numbers, colored according to the $A$ value they correspond to. At the bottom of figure~\ref{fig:AoAEffect}, in panels (a1)--(d1) we plot $y_{1/4}(t)=y(-1/2,t)$ for the same time range as for the color plots and the same color code as above.
These are another guide to the overall dynamics in the large-amplitude steady state. We use $y_{1/4}(t)$ specifically because it distinguishes membranes that flipped from those that did not: generally $y_{1/4}<0$ corresponds to flipped membranes and $y_{1/4}>0$ to non-flipped membranes.

Figures~\ref{fig:AoAEffect}(a) and (b) show examples of membranes that flip with any of three values of $A$ (increasing from left to right). The membrane parameters in (a) and (b) are identical: $R_1=10^1$, $R_3=10^2$, and $T_0=10^0$. However, the kinematic parameters of the tacking maneuver differ: row (a) is with $\Delta t=10$ and $\xi=-0.2$  whereas row (b) is with $\Delta t=80$ and $\xi=0.2$. We typically find that it takes less time for membranes to flip when the tacking maneuver is performed with a larger $A$. A possible explanation is that a larger $A$ 
implies larger pressure forces on the membrane, which can result in them flipping more easily.
In particular, for panels (a) and (b) we see that $t_{i}^{\mathrm{flip}}\{20\degre\}< t_{i}^{\mathrm{flip}}\{10\degre\}<t_{i}^{\mathrm{flip}}\{5\degre\}$ for $i=1,2,3$.

The approximate flip-times and the time evolution of the membrane oscillations can be visualized through the $y_{1/4}(t)$ plots at the bottom of figure~\ref{fig:AoAEffect}. In (a1) and (b1) the green $y_{1/4}(t)$ curves (with $A=20\degre$), are the first ones to satisfy that $\Upsilon$ in \eqref{eq:flipTimeRatio} is less than 0.01 (and 0.001)---the criterion for $t_2^{\mathrm{flip}}$ (and $t_3^{\mathrm{flip}}$), which can be roughly seen through the oscillations decaying to small magnitudes earlier in time. The next $y_{1/4}(t)$ curve to satisfy these criteria is the purple one ($A=10\degre$) and the last one is the orange with $A=5\degre$. Although the oscillation amplitudes for the orange curves are eventually smaller than the others, this difference is too small to be captured by our definitions of flip times. 

In figures~\ref{fig:AoAEffect}(c) and (d) we demonstrate two examples of more `unusual' motions, where (by at least one of the flip-time measures) the time it takes for a membrane to flip is shorter when the tacking maneuver is performed with $A=10\degre$ versus $20\degre$; with $A=5\degre$ the membranes did not flip (DNF). 
Compared to figures~\ref{fig:AoAEffect}(a) and (b), these two cases have a smaller $R_3$ ($10^1$) but the same $T_0$ ($10^0$).
In figures~\ref{fig:AoAEffect}(c1) and (d1) the orange $y_{1/4}(t)$ curves lie above the dashed zero-line since the membranes did not flip with $A=5\degre$. The green $y_{1/4}(t)$ curves (with $A=20\degre$) are the first ones to drop below the zero-line which lead to smaller $t_1^{\mathrm{flip}}$ compared to the purple curves (with $A=10\degre$).
Although the first measure of flip time still satisfies the expected inequality given the value of $A$, i.e.\ $t_1^{\mathrm{flip}}\{20\degre\}< t_1^{\mathrm{flip}}\{10\degre\}$, $t_{2}^{\mathrm{flip}}$ and $t_{3}^{\mathrm{flip}}$ satisfy the reverse inequalities, as highlighted in boldface text above panels (c1) and (d1). In the lists of flip times in figure~\ref{fig:AoAEffect}, we use $\geq 500$ if $t_{i}^{\mathrm{flip}}\geq 500$, where 500 is the final time of the simulation.

In figure~\ref{fig:effectOfR3andT0} we demonstrated the effect of $R_3$ and $T_0$ on the membrane dynamics by fixing $R_1$, $A$ (same three values as in figure~\ref{fig:AoAEffect}) and other kinematic parameters. We found that membranes always flip when $A=20\degre$. For smaller values of $A$ (i.e.\ $5\degre$ and 10$\degre$) the relative magnitudes of $R_3$ and $T_0$ had a greater effect on determining whether a membrane flips or not after the tacking maneuver and our choice of parameters in figure~\ref{fig:AoAEffect} also shows this. It is possible that the non-flipped state is less stable than the flipped state, and may not be an equilibrium (stable or unstable) for a large enough angle of attack, such as $20\degre$.

In summary, we find that the larger the value of $A$ the more likely it is for membranes to flip.  Generally, larger values of $A$ also lead to shorter flip times. 
However, there are a few exceptions (10 out of the 299 cases that flipped, i.e. the cases in groups I and J) that give shorter flip times with smaller $A$. These cases mostly occurred with $R_1\geq 10^1$, $\xi=0.2$, and $\Delta t =10$. These values of the dynamical parameters can give a larger acceleration to the fluid-sail system during the tacking maneuver, and lead to more persistent oscillations when~$A = 20\degre$.

\subsection{Effect of tacking time}\label{sec:tacktime}

\begin{figure}[H]
    \centering
    \includegraphics[width=\textwidth]{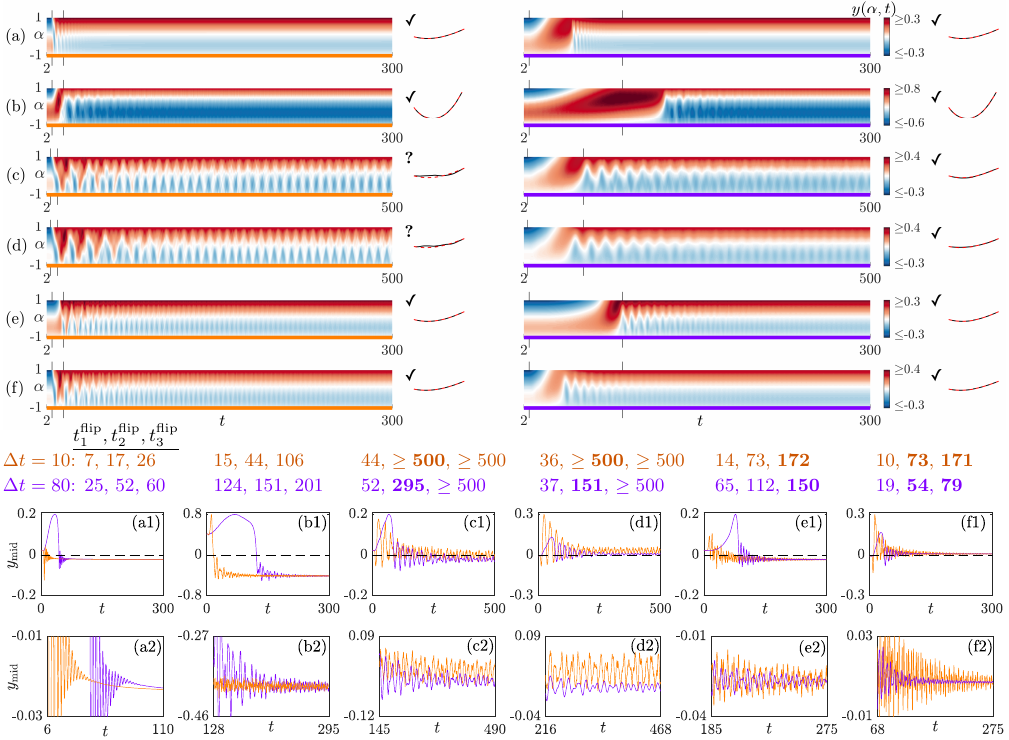}
    \caption{(Effect of $\Delta t$.) The color plots show $y(\alpha,t)$, over selected time intervals that illustrate the main dynamics.  Each column corresponds to a different value of tacking time, $\Delta t$: 10 (left) and 80 (right); distinguished by the orange and purple horizontal time-axes. The rest of the parameters $\{R_1,R_3,T_0,A,\xi\}$ are: (a) $\{10^{-2},10^2,10^{-1},10\degre,0.2\}$, (b) $\{10^{1},10^1,10^{-2},20\degre,0\}$, (c) $\{10^{2},10^2,10^{-2},10\degre,0\}$, (d) $\{10^{2},10^2,10^{0},10\degre,0\}$, 
    (e) $\{10^{1},10^2,10^{-2},10\degre,-0.2\}$, and (f) $\{10^{1},10^2,10^{0},10\degre,0.2\}$. Below the color plots, we give the computed flip times $t_1^{\mathrm{flip}}$, $t_2^{\mathrm{flip}}$, and $t_3^{\mathrm{flip}}$ as lists of three numbers. In panels (a1)--(f1) $y_{\mathrm{mid}}(t)$ is plotted, with the corresponding color for each $\Delta t$ and with the same time range as for the color plots. The black dashed lines show the zero line. At the bottom portion of the figure, in panels (a2)--(f2), we show a zoomed-in version of $y_{\mathrm{mid}}$ to illustrate more clearly which $\Delta t$ has shorter flip times.}
    \label{fig:tacktimeEffect}
\end{figure}

Another quantity that can be controlled in the tacking maneuver is the total tacking time $\Delta t$. In figure~\ref{fig:tacktimeEffect} we show typical examples of how membrane motions are affected by $\Delta t$ using two values: 10 and 80; represented by the orange and purple lines, respectively. At the top of figure~\ref{fig:tacktimeEffect} (in panels (a)--(f)) color plots are used to visualize $y(\alpha, t)$. Below the color plots, we give $t_1^{\mathrm{flip}}$, $t_2^{\mathrm{flip}}$, and $t_3^{\mathrm{flip}}$ as lists of three numbers, colored according to the $\Delta t$ value they correspond to. In panels (a1)--(f1) we now plot $y_{\mathrm{mid}}(t)=y(0,t)$ in each case with the same time interval as in panels (a)--(f). This allows us to better visualize the motions near the large-amplitude steady state. In figures~\ref{fig:AoAEffect}(a1)--(d1) we plotted instead $y_{1/4}(t)$, because in all of those cases, $y_{\mathrm{mid}}(t)$ was close to zero, so $y_{\mathrm{mid}}(t)$ did not show which membranes flipped and which did not. At the bottom portion of figure~\ref{fig:tacktimeEffect}, in panels (a2)--(f2), we zoom in on $y_{\mathrm{mid}}$ in certain time ranges to see more clearly which tacking time gives shorter flip times.

The flip times are generally shorter with $\Delta t=10$, since the tack is performed over a shorter time interval. We show examples of this typical case in figures~\ref{fig:tacktimeEffect}(a) and (b), where the 
small oscillations during and after the tacking maneuver fade away earlier in the left column than in the right column. This can be seen from the fading zig-zag patterns in the color plots but also more clearly in panels (a2) and (b2), where the $y_{\mathrm{mid}}$ oscillations decay faster for the orange lines ($\Delta t=10$) than for the purple lines ($\Delta t=80$). These two cases have significantly different membrane parameters: in (a) the mass is small ($R_1=10^{-2}$), the stretching rigidity is large ($R_3=10^2$), and the pretension is moderately small ($T_0=10^{-1}$) whereas in (b) $R_1$ is large ($10^{-1}$), $R_3$ is smaller ($10^1$) but still moderately large, and $T_0$ is small ($10^{-2}$). The other two kinematic parameters are also different: $A=10\degre$ and $\xi=0.2$ in (a), and $A=20\degre$ and $\xi=0$ in (b). This demonstrates that shorter flip times can be achieved with a small $\Delta t$ in different regimes of the parameter space, spanning different magnitudes of the remaining five parameters.


We show two special cases at large mass ($R_1=10^2$) in figures~\ref{fig:tacktimeEffect}(c) and (d), where the membranes start to oscillate after the tacking maneuver, 
with a traveling wave motion that persists for the remaining simulation time. In (c) and (d)---as for all other cases in figure~\ref{fig:tacktimeEffect}---$t_1^{\mathrm{flip}}\{10\}<t_1^{\mathrm{flip}}\{80\}$ since the membrane approaches the flipped shape faster with a smaller tack time $\Delta t$. This type of motion implies that the membrane shape never becomes very close to the flipped shape, which results in $t_2^{\mathrm{flip}},t_{3}^{\mathrm{flip}}\geq 500$. With $\Delta t=80$ (column 2), the oscillations after the tacking maneuver are also there, and again persist in time, but the oscillation amplitudes in this case are smaller, giving $t_{2}^{\mathrm{flip}}\{80\}<500\leq t_{2}^{\mathrm{flip}}\{10\}$. 

In panels (e) and (f) we show two additional examples of unusual behaviors that happen with a smaller mass ($R_1=10^1$), where at least one of the flip-time measures is shorter for $\Delta t=80$ than for 10.
Compared to (c)--(d), the oscillation amplitudes in (e) and (f) are much smaller and the oscillations seem to decay in time. This can be seen from the intensity of the colors in the color plots and from the significantly smaller scales in the vertical axes of the zoomed-in $y_{\mathrm{mid}}$ plots in panels (e2) and (f2). 

In this section, we have shown that a smaller tack time, $\Delta t=10$ rather than 80, generally leads to shorter flip times. The exceptions of larger flip times with $\Delta t=10$ (shown in figure~\ref{fig:tacktimeEffect}) are very rare---14 of the 299 cases that flipped (i.e. the cases in groups I and J in figure~\ref{fig:noflipping}). These cases occur with all $\xi$ values, $A=5\degre$ or $10\degre$, and have $R_1\geq 10^1$. Such exceptions may occur because $\Delta t=10$ gives a larger acceleration to the fluid-body system which can sometimes persist for long times. Whether the membrane flips or not is affected by $\Delta t$ in a small number of cases. In 11 cases, all with $R_1=10^1$ or $10^2$ and $A=5\degre$ or $10\degre$, the membrane flipped with $\Delta t = 10 $ but not with $\Delta t = 80$. One case ($R_1 = 10^2$) flipped with $\Delta t$ = 80 but with $\Delta t = 10$ it instead oscillated for long times.

\subsection{Effect of angle-of-attack transition kinematics (profile)}\label{sec:offset}

Another main feature of the angle-of-attack kinematics that we can control during the tacking maneuver is the concavity of $\Theta(t)$. The sign and magnitude of the ``offset'' $\xi$ specify the exact profile of $\Theta(t)$: $\xi>0$ gives a concave down profile, $\xi=0$ gives nearly zero concavity, and $\xi<0$ gives a concave up profile. It is desirable to achieve shorter flip times and in our model we might expect this to occur when $\xi$ is large. This is because a concave down $\Theta(t)$ implies that the angle of attack tends to $A$ sooner during the tacking maneuver.

\begin{figure}[H]
    \centering
     \includegraphics[width=\textwidth]{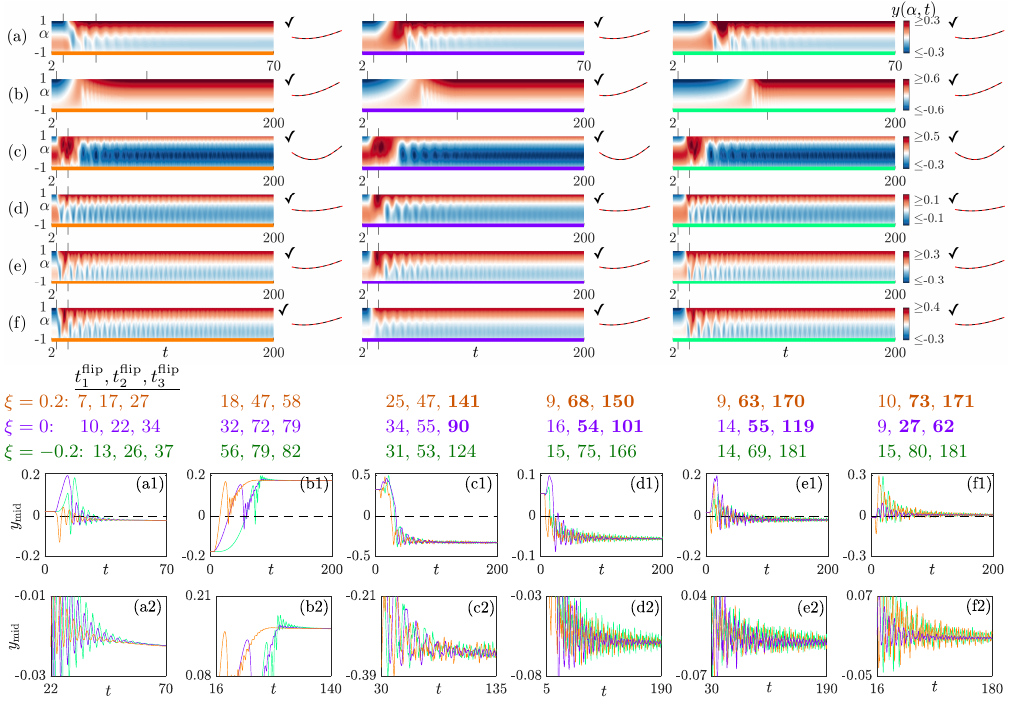}
    \caption{(Effect of $\xi$.) The color plots, lists of numbers, and the bottom portions of the figure, show the same quantities as in figure~\ref{fig:tacktimeEffect}. Each column corresponds to a different value of offset, $\xi$: $0.2$ (left), $0$ (middle), and $-0.2$ (right); distinguished by the orange, purple, and green horizontal time-axes. The rest of the parameters $\{R_1,R_3,T_0,A,\Delta t\}$ are: (a) $\{10^{-1},10^2,10^{-1},10\degre,10\}$, (b) $\{10^{0},10^2,10^{0},20\degre,80\}$, (c) $\{10^{1},10^1,10^{0},10\degre,10\}$, (d) $\{10^{1},10^2,10^{0},5\degre,10\}$, 
    (e) $\{10^{1},10^2,10^{-1},10\degre,10\}$, and (f) $\{10^{1},10^2,10^{0},10\degre,10\}$. }
    \label{fig:offsetEffect}
\end{figure}

We use three angle-of-attack profiles $\Theta(t)$ that are determined by different values of the offset parameter $\xi$ (chosen as $0.2$, 0, and $-0.2$, shown in figure~\ref{fig:AoAschem}(c)). In figures~\ref{fig:offsetEffect}(a) and~(b) we show typical membrane motions: $\xi = 0.2$ gives the shortest flip time, then 0, and finally $-0.2$. 
In figure~\ref{fig:offsetEffect}(b) the tacking maneuver is performed over a longer time interval than in (a): $\Delta t=80$ versus 10. In this case, the offset value $\xi$ has a larger effect on $t_1^{\mathrm{flip}}$.
Although the flip time is expected to be smaller for larger values of $\xi$, this is not always the case. In figures~\ref{fig:offsetEffect}(c)--(f) we display four examples of this special nonmonotonic behavior, where $\xi=0$ (middle column) results in membranes flipping faster than when $\xi=0.2$ (left column).

Usually in sailing competitions a proper tack would start with a relatively slow and smooth change in the angle of attack to preserve momentum and allow the sailboat to move upwind. As the sailboat comes head-on to the wind and speed is lost, then the sail membrane should be turned more quickly to complete the tack~\cite{northsails,bethwaite2013fast}. 
This strategic maneuver would be achieved with a negative $\xi$, or equivalently a concave up $\Theta(t)$. In most cases we find that larger $\xi$ leads to shorter flip times but in our model the sailboat is moving at a constant speed. This is a simple model that allows us to characterize the sail dynamics during tacking maneuvers for a six-dimensional parameter space, but future work could include incorporating this sailboat momentum loss by making the oncoming flow speed (the wind speed relative to the boat) vary to reflect the changing speed of the boat.










We have already seen multiple instances where $A=5\degre$ and $10\degre$ in combination with specific values of the other parameters lead to membranes that do not flip. In these cases, to enable sail flipping we may transiently increase the sail angle of attack above the final value by some amount. This amount is approximately given by an \textit{overshoot} parameter that defines the maximum value of $\Theta(t)$ that the sail attains during the tacking maneuver, which will be greater than the final value $A$.
For the remainder of this section we focus on such non-flipped membranes and add an overshoot parameter $\sigma$ as shown in figure~\ref{fig:AoAschem}(d), and see what the differences in their dynamics are. This $\sigma$ sets the transient increase in the maximum angle of attack above $A$ during tacking. A similar parameter was used in the optimization of tacking for an actual racing sailboat~\cite{hansen2019maneuver}.


\begin{figure}[H]
    \centering
    \includegraphics[width=\textwidth]{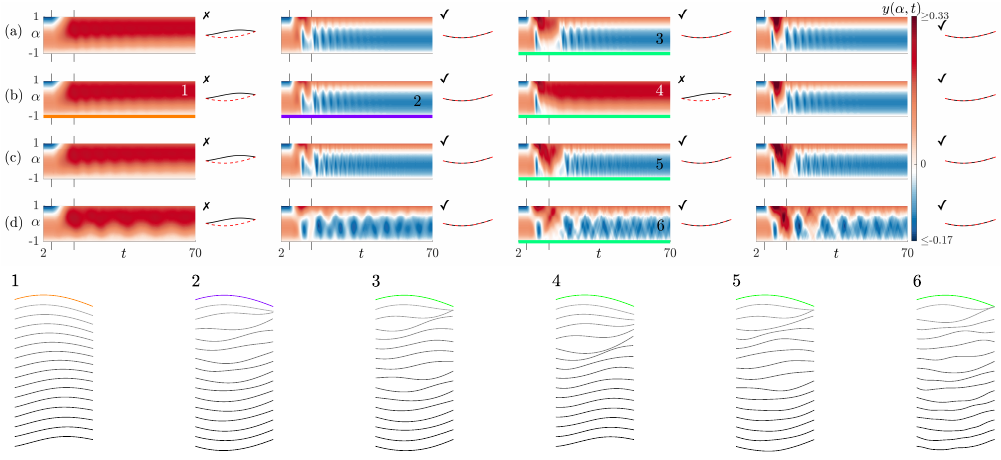}
    \caption{(Effect of $\sigma$.) The color plots show the membrane deflection $y(\alpha,t)$ with $\alpha\in[-1,1]$ and $t\in[2,70]$ in the vertical and horizontal axis respectively. We fix $\{R_3,T_0,A,\Delta t,\xi\}=\{10^2,10^{-2},5\degre,10,0\}$. Each row has a different value of $R_1$: (a) $10^{-2}$, (b) $10^{-1}$, (c) $10^0$, and (d) $10^1$.  Each column has a different value of the overshoot parameter, $\sigma$: $0$ (first)---no overshoot, $0.2$ (second), $0.4$ (third), and $0.6$ (fourth). At the bottom of the figure we show six examples of membrane snapshots arrayed vertically moving downward with increasing time. The numbers at the top left of each sequence of snapshots indicate the $(R_1,\sigma)$ combination that has the corresponding color plot associated with it; 1.\ $(10^{-1},0)$, 2.\ $(10^{-1},1.2)$, 3--6.\ with $R_1\in[10^{-2},10^1]$ and $\sigma=0.4$. The motions in 4 and 5 are shown in the supplementary movies named ``Fig14\textunderscore 4\dots mp4'' and ``Fig14\textunderscore 5\dots mp4,'' respectively.}
    \label{fig:overshootT0minus2}
\end{figure}

In figure~\ref{fig:overshootT0minus2} we consider examples of membranes that with $\sigma=0$ do not flip (column 1) but generally flip with $\sigma\geq 0.2$. 
We fix the membrane parameters as: $R_3=10^{2}$ and $T_0=10^{-2}$, 
and vary $R_1$ in the color plots of membrane deflection from $10^{-2}$ to $10^1$, from top to bottom. The other kinematic parameters are fixed as $A=5\degre$, $\Delta t=10$, and $\xi=0$.
Each column of color plots has a different value of $\sigma$: 0 (first), 0.2 (second), 0.4 (third), and 0.6 (fourth). The reddish colors throughout $\alpha\in[-1,1]$ indicate that the membrane deflection satisfies $y(\alpha,t)\geq 0$ for all $t\geq t_i+\Delta t$ which is equivalent to  membranes not flipping. The ones with small $R_1$ ($\leq 10^0$) in panels (a)--(c) become almost uniform in color across $t$ as $t\to 70$ showing  
that the membranes eventually after the tack attain a steady single-hump shape. In the second column, the same sail performs the tack with an overshoot of $\sigma=0.2$. In this case, for any of the $R_1$ values considered here (panels (a)--(d)) the membrane flips but during the tacking maneuver (and for some time interval after the tack is complete) it undergoes a series of small oscillations about the single-hump shape. In the third column the overshoot parameter is increased to $\sigma=0.4$. This generally has a similar effect to $\sigma=0.2$ but, surprisingly, in this scenario the membrane with $R_1=10^{-1}$ (example labeled with the number~4 in (b)) initially flips but soon after returns to a non-flipped configuration which eventually becomes steady at later times. This sudden shape transition from non-flipped to flipped and back to non-flipped, is clearly seen in the vertically arrayed membrane snapshots in example 4 at the bottom of figure~\ref{fig:overshootT0minus2}. 
With $\sigma=0.4$ the membranes that flipped (examples labeled~3, 5, and 6) have larger flip-times than those with $\sigma=0.2$ (second column).
Additionally, in panel (d) at the largest $R_1$ here ($10^{1}$), the membrane example labeled~6, starts to oscillate with a traveling wave motion shown by diagonal bands, unlike $\sigma=0.2$. 

In the last column the overshoot parameter is further increased to $\sigma=0.6$. Similarly to $0.2$, now membranes flip for any of the $R_1$ values ((a)--(d)) presented in figure~\ref{fig:overshootT0minus2}. The dynamics share some similarities but also exhibit some differences. For instance, for $R_1\in[10^{-2},10^0]$ (panels (a)--(c)) the dynamics in columns 2 and~4 are almost the same after the tack.  Perhaps the most distinct feature between $\sigma=0.2$ and 0.6 is that in the latter case the membrane during the tack attains larger deflection amplitudes (very dark red regions in the time interval $t\in [t_i,t_i+\Delta t]$). For $R_1=10^1$ (panel (d)) the larger overshoot results in membranes that oscillate between the flipped and non-flipped shapes for longer. This is seen through the reddish regions of $y(\alpha,t)$ spanning larger areas of $\alpha\in[-1,1]$ after the tack in the fourth column. Apart from this large-mass case, the remaining cases have flip-times similar to those with $\sigma=0.4$.

\begin{figure}[H]
    \centering
    \includegraphics[width=\textwidth]{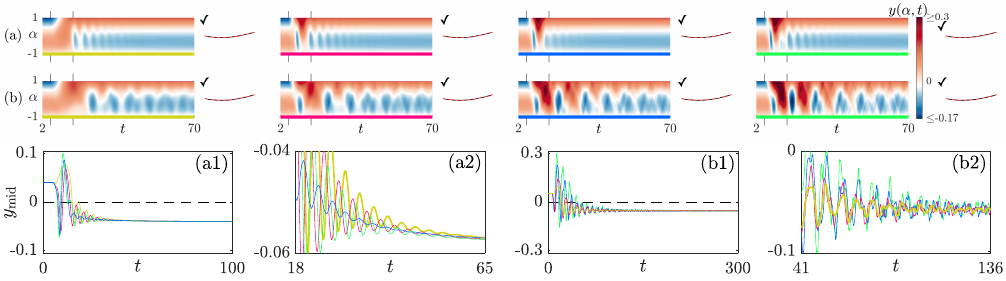}
    \caption{(Effect of $\sigma$.) The color plots at the top show $y(\alpha,t)$, same as figure~\ref{fig:overshootT0minus2} but with $T_0=10^0$, (a) $R_1=10^{-1}$ and (b) $R_1=10^1$ (and all other parameters the same). Each column has a different value of $\sigma$: 0 (first), 0.2 (second), 0.4 (third), and 0.6 (fourth); distinguished by the gold, magenta, blue, and green horizontal time-axes. In contrast to figure~\ref{fig:overshootT0minus2}  membranes now flip even when $\sigma=0$ (first column). Below the color plots of membrane deflection, we plot in (a1) and (b1) $y_{\mathrm{mid}}(t)$, with the corresponding color for each $\sigma$ and with the same time range as for the color plots. The black dashed lines show the zero line. Next to each of these panels, in (a2) and (b2), we show a zoomed-in version of $y_{\mathrm{mid}}$ to illustrate more clearly which $\sigma$ leads to shorter flip times. 
    }
    \label{fig:overshootT00}
\end{figure}
In figure~\ref{fig:overshootT00} we use $T_0=10^0$ which is a larger pretension value compared to the one in figure~\ref{fig:overshootT0minus2}, and use $R_1$ equal to (a) $10^{-1}$ and (b) $10^1$, while keeping the rest of the parameters the same as in figure~\ref{fig:overshootT0minus2}. These choices result in membranes that flip even without the overshoot parameter (i.e.\ $\sigma=0$ in the first column of figures~\ref{fig:overshootT00}(a) and (b)). Panels (a1)--(a2) indicate that the membrane with $\sigma=0.4$ (blue line) flips earliest and has the fastest decay of membrane oscillations. For panel (a), $t_2^{\mathrm{flip}}$ and $t_3^{\mathrm{flip}}$ are indeed smallest with $\sigma=0.4$, but $t_1^{\mathrm{flip}}$ is 5\% smaller for $\sigma=0.2$.
Panels (b1)--(b2) indicate that the membrane without overshoot ($\sigma=0$) flips earliest, and this is the case according to all three flip time measures. 
To summarize, we find that a concave down angle-of-attack profile ($\xi>0$) results in membranes that are more likely to flip. A possible reason is that the membrane experiences a larger pressure jump when the angle of attack tends to $A$ sooner during the tacking maneuver. For the sails that flip, we find that a larger value of $\xi$ usually yields shorter flip times. There are a few exceptions of larger flip times with $\xi>0$ (e.g. those shown in figure~\ref{fig:offsetEffect}). In such cases, $\xi=0.2$ may give a larger acceleration to the fluid-sail system that can persist for long times, similarly to the cases discussed in the previous section with $\Delta t=10$ instead of 80. The exceptions for $\xi=0.2$ generally occur at small tack time ($\Delta t=10$) and large $T_0$ ($10^0$).

We find that larger $\sigma$ generally makes membranes more likely to flip, as it corresponds to a larger maximum angle of attack, and thus larger pressure forces.

\section{Slack sail}\label{sec:slack}

Real sails are usually not completely flat even in the absence of wind. They tend to be wrinkled when they are not pulled taut and their pretension is zero or almost zero. In such cases the ends may be closer together than the sail length~\cite{sneyd1984aerodynamic}. 
We incorporate this in our model and simulations by bringing the sail's  ends (leading and trailing edges) inward towards each other which effectively corresponds to reducing the membrane's tension to zero or negative values before the tacking maneuver. The tension is given by
$T = T_0+R_3(\partial_\alpha s-1)$. Bringing the ends toward each other reduces $\partial_\alpha s$ below 1 which can make $T$ negative depending on $T_0$, $R_3$, and $\partial_\alpha s$. When the tension is negative, the membrane would buckle into a nonflat (wrinkled) rest state with zero tension---a ``slack sail"---in the absence of fluid forces (as there is no bending rigidity to prevent buckling). However, with a nonzero background oncoming flow, fluid pressure forces pull the membrane into a curved shape with positive, but reduced, tension when the ends are moved towards each other.
We now show how chord-shortening affects the dynamics of flipped membranes through examples.  

In this new configuration, the time-dependent boundary conditions at the leading edge remain the same as in~\eqref{eq:bc_LE} (i.e.\ $ x(-1,t) = -1$ and $y(-1,t) = 0$) but the membrane's trailing edge~\eqref{eq:bc_TE} moves toward the leading edge, i.e. the chord length decreases:
\begin{equation}\label{eq:slackBC}
     x(1,t) -x(-1,t) = (\text{chord length})\times \cos (-A)\quad ;\quad y(1,t) -y(-1,t) = (\text{chord length})\times \sin (-A),
\end{equation}
where 
\begin{equation}\label{eq:slackChord}
    \text{chord length}=
     \begin{cases}
     2,& 0<t\leq t_I, \qquad \hfill \leftarrow\text{rest}\\
     2+\displaystyle\frac{(C-2)}{t_s}(t-t_I), & t_I<t\leq t_I+t_s, \qquad \hfill \leftarrow\text{chord-shortening}\\
     C,& t_I+t_s\leq t<t_I+t_s+1.\qquad  \hfill \leftarrow \text{rest}
     \end{cases} 
\end{equation}
Here $t_I$ is the initial time before the chord-shortening begins, $t_s$ is the total length of time over which the chord-shortening takes place (done slowly here, over 10 time units), and $C$ is the final, user-prescribed distance between the membrane's ends. During the chord-shortening, the distance between the two ends changes linearly from 2 to $C$ over $t_s$ time units. 

The tacking maneuver is performed when $t>t_I+t_s+1$ and during that time the sail membrane's time-dependent boundary conditions become:
\begin{align}
\text{leading edge:}\quad&
    x(-1,t) = -1; \quad y(-1,t) = 0,\label{eq:bc_LE_endshort}\\
\text{trailing edge:}\quad&  x(1,t) = x(-1,t) + C\cos(\Theta(t-(t_I +t_s + 1))); 
\quad & y(1,t) =  C\sin(\Theta(t-(t_I +t_s + 1))).\label{eq:bc_TE_endshort}
\end{align}







We focus on membranes that flip without chord-shortening. We use 
as examples membranes with two values of $R_1$ ($10^{-1}$ and $10^0$), two values of $T_0$ ($10^{-2}$ and $10^{-1}$), and $\{R_3,A,\Delta t,\xi\}=\{10^{2},20\degre,10,0\}$.

\begin{figure}[H]
    \centering
    \includegraphics[width=\textwidth]{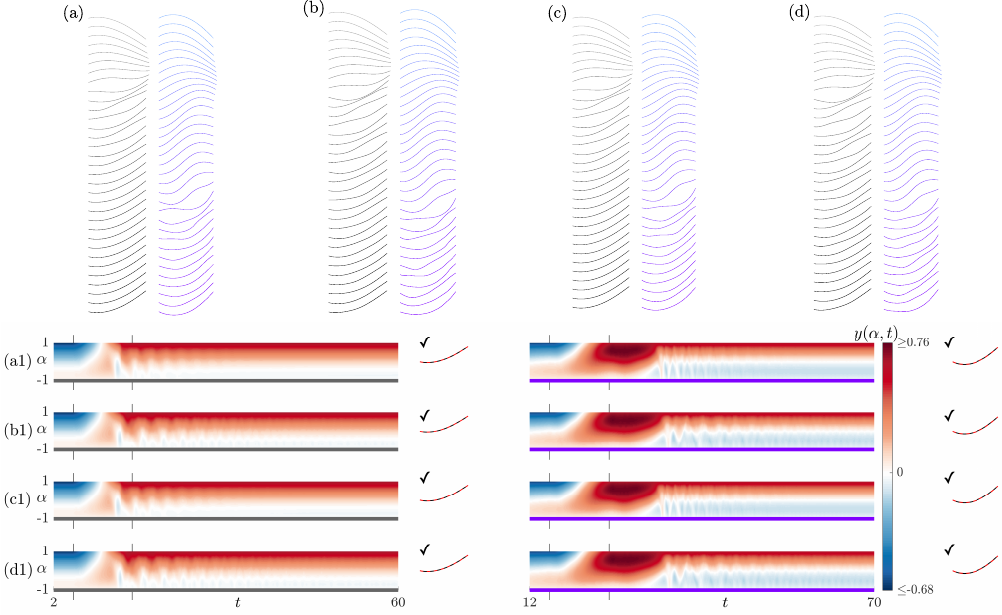}
    \caption{(Effect of chord-shortening.) Examples of membrane dynamics during the tacking maneuver for membranes without chord-shortening and with chord-shortening from 2 to $C=1.9$. Panels (a)–(d) show sequences of membrane snapshots arrayed vertically from top to bottom with increasing time, in darkening shades from gray to black (on the left column of each panel, without chord-shortening), and blue to purple (on the right column of each panel, with chord-shortening). The snapshots are in the time interval $[t_i,t_i+4\Delta t]$. The membrane parameters are: $(R_1,T_0)$ equal to (a) $(10^0,10^{-1})$, (b) $(10^1,10^{-1})$, (c) $(10^0,10^{-2})$, (d) $(10^1,10^{-2})$, and $R_3=10^2$ in all cases.
    The kinematic parameters are: $\{A,\Delta t,\xi\}=\{20\degre,10,0\}$. The color plots in panels (a1)--(d1) show $y(\alpha,t)$ for the case without chord-shortening in the first column (gray horizontal time-axis) and with the chord-shortening in the second column (purple horizontal time-axis).}
    \label{fig:slack}
\end{figure}

We present in figure~\ref{fig:slack} a comparison between the sail membrane dynamics  without and with chord-shortening (with $C=1.9$ in~\eqref{eq:slackChord}). In panels (a)--(d) we show sequences of membrane snapshots arrayed vertically from top to bottom with increasing time, in darkening shades from gray to black for cases without chord-shortening and from blue to purple for cases with chord-shortening. The snapshots reveal that with the chord-shortening (purple) the membrane is generally more curved. Larger deflections can be seen through the darker shades of red in the right color plots in figures~\ref{fig:slack}(a1)--(d1), which correspond to the upper ranges of $y(\alpha,t)$ in the color bar. As we have already discussed in \S\ref{sec:tacking}, higher curvature usually implies that the membrane has more difficulty flipping. This is evident through both the vertically arrayed (purple) snapshots and the color plots on the right-hand side of (a1)--(d1). In particular, in figures~\ref{fig:slack}(a)--(d) we see that the slack membrane adopts an almost steady shape with an inflection point closer to the leading edge towards the end of the tacking maneuver; a shape that it maintains for a time duration $\approx \Delta t$ after the tacking maneuver. This corresponds to the dark-red region on the right hand-side of panels (a1)--(d1), which covers a large time interval mostly at $t_i+\Delta t < t \lesssim t_i+2\Delta t$ (after the right set of hash marks).  Later, the slack membrane starts to oscillate, before eventually flipping and settling down to a steady single-hump shape.

Using the same parameters, but bringing the sail ends closer together, with $C=1.5$ in \eqref{eq:slackChord}, results in membranes that do not flip in any case and so we omit their snapshots here. 

\section{Conclusions}\label{sec:conclusions}

We have studied the dynamics of sail membranes during tacking maneuvers using a membrane--vortex-sheet model. We analyzed how the dynamics change based on seven main parameters that can be divided into two groups: those that determine the properties of the sail and those that control the kinematics of the tacking maneuver. 
We found that the relative magnitudes of the steady parameters---stretching rigidity $R_3$, pretension~$T_0$, and angle of attack $A$---play an important role in whether a sail flips to the mirror-image configuration of its initial shape or not, and that the dynamical parameters---membrane mass density $R_1$, tack time $\Delta t$, offset~$\xi$---mainly affect the flip times.

For steady membrane shapes at different angles of attack to the wind, we found that increasing the angle of attack (for a fixed pretension) results in sail membranes with larger deflections and higher curvatures closer to the leading edge. The point of maximum curvature of the sail was close to a local minimum of the pressure jump. Using the steady parameters $R_3$, $T_0$, and $A$ we obtained 30 steady membrane shapes that we then used as the initial membrane shapes for 540 unsteady tacking simulations. We found that membranes either flipped or did not flip, with a few indeterminate cases. Using a metric that describes the distance of the time- and spaced-averaged membrane shape from the flipped state we ranked the membranes along a continuum from flipping to not flipping. This showed that the computed membrane motions could be sorted into ten groups, each with similar steady-state shapes, and that groups of membranes that did not flip were associated with particular values of $R_3$, $T_0$, and $A$.
Membranes with $A=20\degre$ always flipped. 
For smaller values of $A$, 
either one or both of $R_3$ and $T_0$ had to be large for membranes to flip after tacking. This is because large $R_3$ and $T_0$ both resulted in smaller membrane deflections, making flipping more likely. 

We found that the membrane mass parameter $R_1$ only determines whether flipping occurs in a small number of cases, where flipping did not occur at small $R_1$ but did occur at larger $R_1$.
The clearest effect of the membrane mass parameter 
$R_1$ was to decrease the frequency of the oscillations at large $R_1$, where the membrane inertia term strongly resists high-frequency motions. Large $R_1$ also resulted in membranes that took longer to converge to a steady state. In a few cases, with the largest $R_1$ ($10^2$), membranes had sustained oscillations until the end of the simulation, and flipping was therefore indeterminate. At small $R_1$, the membrane mass was negligible compared to the fluid mass and so the results at $R_1<10^0$ were similar to those with $R_1=10^0$. 

When the other two membrane parameters, $R_3$ and $T_0$, were large the membranes had smaller membrane deflections and flipped. If instead these two parameters were small, the final angle of attack~$A$ had to be large for the membranes to flip.

For the kinematic parameters we also considered how quickly the membranes flip (that is, if they flip) using different ``flip time" measures. We found that it typically takes less time for membranes to flip when the tacking maneuver is performed with a larger $A$. This is because a larger $A$ gives larger pressure forces on the membrane which can result in them flipping more easily. However, we also observed examples of unusual motions that revealed that certain combinations of the remaining five parameters can lead to membranes that flip faster when the tacking maneuver is performed with $A=10\degre$ compared to $20\degre$. Another important quantity that we investigated was the total tacking time $\Delta t$. We showed that the flip times are generally shorter with small $\Delta t$ (10 versus 80). This is expected because the tack is performed over a shorter time interval. However, as with $A$, there were special cases for which, by at least one of the flip-time measures, the membrane approached the flipped shape faster with $\Delta t=80$ than with 10. Such cases corresponded to membranes that started to oscillate after the tacking maneuver, with a traveling wave motion that persisted until the end of the simulation time. Larger $\Delta t$ gave a shorter flip time because the oscillation amplitudes in these cases were smaller.

Another main feature of the angle-of-attack kinematics that we varied was the concavity of the angle-of-attack profile. The majority of sails (given combinations of the other parameters) achieved shorter flip times with concave down angle-of-attack profiles (large offset values $\xi$) since then the angle of attack approached its final value sooner. 
Exceptions of longer flip times with large $\xi$ were rare but did occur for $\Delta t=10$. In these cases a large $\xi$ gave a larger acceleration to the fluid-sail system that could persist for long times.

For cases that did not flip, we introduced an overshoot parameter $\sigma$ that corresponded to increasing the maximum value of angle of attack above the final value $A$. This overshoot parameter, in all but one example, caused the non-flipped membranes to flip. 
In such cases, increasing the angle of attack above the final angle yields pressure forces that are large enough to cause the sail to flip. Generally, larger overshoot parameters resulted in larger membrane deflections during the tacking maneuver, and in oscillatory motions that switched between the flipped and non-flipped shapes for longer. Therefore, to achieve shorter flip times $\sigma$ should not be too large.

Finally, we considered an alternative configuration---a ``slack'' sail---to mimic the fact that real sails are usually not completely flat even in the absence of wind. To do this, we brought the sail's ends inward towards each other and repeated some of our unsteady tacking simulations. This effectively corresponded to reducing the membrane's tension before the tacking maneuver. A comparison between the sail dynamics with and without the chord-shortening showed that with the chord shortened, membranes bulged outward to a greater degree. During the tacking maneuver these higher-curvature chord-shortened membranes developed an inflection point near the leading edge. After the tacking maneuver, the membranes maintained an almost steady sail shape for about $\Delta t$ time units before starting to oscillate and eventually flipping. In general, we found that chord-shortening makes flipping more difficult.

\section*{Acknowledgements}
\vspace{-.25cm}
\noindent C.M.\ thanks Alessandro Podo and especially Olmo Cerri for valuable discussions regarding the angle-of-attack transition kinematics used during sailing competitions. S.A.\ acknowledges support from the NSF-DMS Applied Mathematics program, award number DMS-2204900.

\section*{Supplementary material}
\vspace{-.25cm}
\noindent Supplementary movies along with a caption list are available at \url{https://drive.google.com/drive/folders/1VrtRAFd1dPGQjDOrRcVYolRuqSABL2P6?usp=sharing}.

\appendix

\section{Details about the angle-of-attack transition kinematics}\label{app:AoAtauAndSigma}

\begin{figure}[H]
    \centering   
\includegraphics[width=\textwidth]{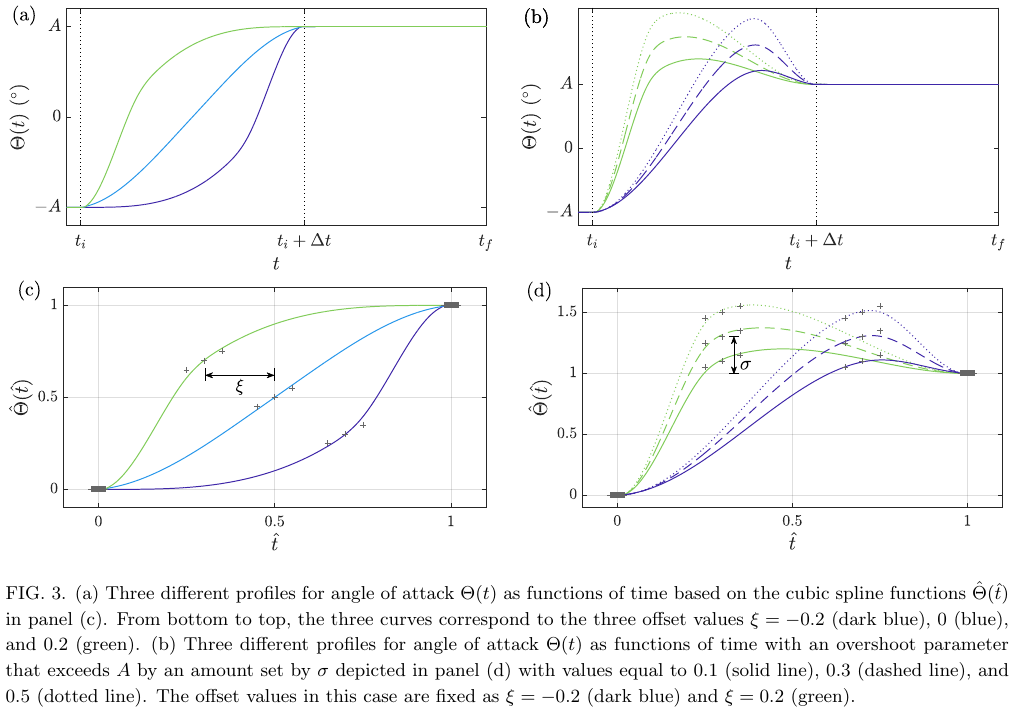}  
\end{figure}

For the angle-of-attack transition kinematics in figure~\ref{fig:AoAschem}(a) we use a cubic smoothing spline function~\cite{reinsch1967smoothing}, computed with the \textsc{Matlab} \texttt{spaps} function~\cite{MATLAB}. The spline function goes through the following points: 
\begin{equation}\label{eq:xLCR}
 \hat{t}_L\in \{-0.02,-0.018,\dots,0.02\},\quad  \hat{t}_{C}\in \{-0.05,0,0.05\},\quad \hat{t}_R\in\{0.98,0.982,\dots,1.02\}, 
\end{equation}
shown as gray crosses in figures~\ref{fig:AoAschem}(c) and (d). The vectors $\hat{t}_L$, $\hat{t}_C$, and $\hat{t}_R$ in~\eqref{eq:xLCR}
are of size $1\times 21$,  $1\times 3$, and $1\times 21$, respectively. 


The first arguments of \texttt{spaps} are the
given data points $(\hat{t}_j,\hat{\Theta}_j)$ specified as
\begin{align}
    \hat{t}_j&\in\{\hat{t}_L,\hat{t}_{C}+(0.5-\xi), \hat{t}_R\}\\ 
    \hat{\Theta}_j&\in\{0,\dots,0,\hat{t}_{C}+(0.5+\xi), 1,\dots,1\},\label{eq:yXi}
\end{align}
where $\xi$ controls the concavity of the angle-of-attack profile. Each vector $\hat{t}_j$ and $\hat{\Theta}_j$ is of size $1\times 45$.
Other arguments of \texttt{spaps} function in \textsc{Matlab} are:
the tolerance ($\mathrm{tol}=3\times 10^{-3}$) which sets a balance between spline smoothness and deviation from the spline target points, a weight function of size $1\times 45$ (here chosen as $w\in\{ 10,\dots,10,1,1,1,10,\dots,10 \}$) that specifies the cost of deviation from the 45 spline target points, and $m$ which specifies the order of the smoothing spline, set to to $2$ here for the cubic smoothing spline. 

The output is a B-form describing the spline as a weighted sum, evaluated with \texttt{fnval} in \textsc{Matlab} to obtain a vector $\hat{\Theta}_B$ of a specified size, e.g.\ $1\times 1000$. We also define a new equispaced vector of the same size as $\hat{\Theta}_B$: $\hat{t}\in\{-0.02,-0.019,\dots,1.02\}$, which we use to  modify the definition of $\hat{\Theta}_B$ as follows:
\begin{align}
\hat{\Theta}(\hat{t})= 
\begin{cases} 
0,&\quad \hat{t}<0,\\
\hat{\Theta}_B\left(1-e^{-(\hat{t}/\delta_B)^3}\right)  + \left(1-\hat{\Theta}_B\right)e^{-\left((1-\hat{t})/\delta_B\right)^3},&\quad 0\leq \hat{t}\leq 1,\\
1,&\quad \hat{t}>1,
    \end{cases}
\end{align}
with $\delta_B=0.02$.
Finally, we use the final angle of attack~$A$ to stretch and shift $\hat{\Theta}(\hat{t})$ and obtain the appropriate form for $\Theta(t)$, going from $-A$ to $A$ as shown in figure~\ref{fig:AoAschem}(a). The vertical transformations are 
\begin{equation}\label{eq:shiftyhatB}
    \Theta(\hat{t})=2A\hat{\Theta}(\hat{t})-A,
\end{equation}
and the horizontal transformation to get to $\Theta(t)$ in figure~\ref{fig:AoAschem}(a) is done through the total tacking time $\Delta t$, using $t=t_i+\hat{t}\Delta t$.

In figure~\ref{fig:AoAschem}(d), we present a modified spline that uses an additional parameter $\sigma$ that in a modified version of~\eqref{eq:yXi} corresponds roughly to the amount by which the spline overshoots the final value:
 \begin{equation}
     \hat{\Theta}_j\in\{ 0,\dots,0, \hat{t}_C+\sigma,1,\dots,1\}.
 \end{equation}

\section{Comparison of lift forces for steady membranes and circular-arc membranes}\label{app:liftCircularArc}
We can approximate the membrane by a circular arc that passes through the two endpoints of the membrane 
and whose curvature $\overline{\kappa}$ matches the membrane shape $y(x)$ best. In particular, we find $\overline{\kappa}$ using 
\begin{equation}
\overline{\kappa}=\argmin\limits_{0<\hat{\kappa}\leq 1}\int_{x(-1)}^{x(1)} \left(y^{\hat{\kappa}}_{\text{arc}}(\tilde{x}) - y(\tilde{x})\right)^2 \d \tilde{x},
\end{equation}
where $y^{\hat{\kappa}}_{\text{arc}}(x)$ is the circular arc shape for a given $\hat{\kappa}$.

\begin{figure}[H]
    \centering
    \includegraphics[width=\textwidth]{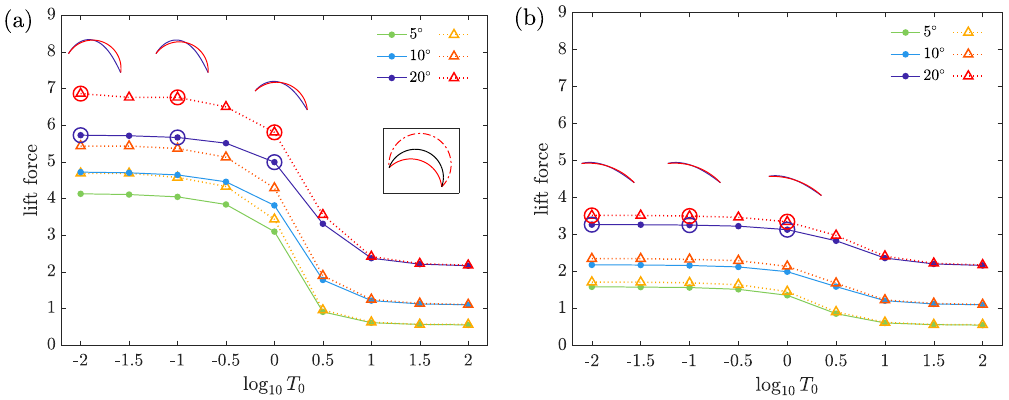}
    \caption{Computed total lift forces (green, light blue, dark blue lines with dot markers) versus $T_0$ for three angles of attack: $5{\degre}, 10{\degre}, 20{\degre}$ and (a) $R_3 = 10^1$, (b) $R_3=10^2$. Superposed are the analytical lift forces given by~\eqref{eq:liftBatchelor} (yellow, orange, red lines with triangles). Comparisons of membranes (dark blue snapshots) versus circular arcs (red snapshots) are shown at $T_0=10^{-2}$, $10^{-1}$, and $10^0$. The dark blue membrane snapshots and the red circular arcs have corresponding~$T_0$ values shown by the circled dark blue dots and the circled red triangles.}
    \label{fig:liftBatchelor}
\end{figure}

When the dimensionless distance between the ends of the membrane (and the circular arc) is 2, the curvature of the circular arc has a minimum value of~0 and a maximum value of 1, which is attained when the circular arc is a semicircle. For all curvature values less than 1, there are two such circular arcs, one more than half of a circle and the other less than half, as shown below the legend in panel (a) of figure~\ref{fig:liftBatchelor} for the case with $T_0 = 10^0$. 
We use the latter because it approximates the membranes better. For this choice, the exact formula for the lift can be derived from results in~\cite{batchelor1967introduction,acheson1990elementary}:
\begin{equation}\label{eq:liftBatchelor}
\mathrm{L}=2\pi\left[\left(\frac{1-\sqrt{1-\overline{\kappa}^2}}{1+\sqrt{1-\overline{\kappa}^2}}\right)^{1/2}
\cos A+\sin A \right],
\end{equation}
where $\overline{\kappa}$ is the curvature of the circular arc.
In figure~\ref{fig:liftBatchelor} we show the total lift force versus $T_0$ for three angles of attack: $5\degre$ (green line with dot markers), $10\degre$ (light blue line with dot markers), and $20\degre$ (dark blue line with dot markers). Panel (a) is for $R_3=10^1$ and panel (b) for $R_3=10^2$. Superposed are the analytical lift forces derived in~\eqref{eq:liftBatchelor} (shown as yellow, orange, and red lines with triangles). We compare the membrane shapes (dark blue snapshots) with the circular arcs (red snapshots) at $T_0=10^{-2}$, $10^{-1}$, and $10^0$. The dark blue membrane snapshots and the red circular arcs have corresponding~$T_0$ values shown by the circled dark blue dots and the circled red triangles. In these examples the circular airfoils have a larger curvature than the membrane shapes (especially in panel (a) with $R_3=10^1$) which explains why the lift forces are significantly overestimated there.

\section{Lift and drag forces}\label{app:liftAndDrag}
We can compute the forces on the membrane at each time using
\begin{equation}\label{eq:force1}
\mathbf{F}(t)=\mathbf{F}_\mathrm{LES}(t)+\int_{-1}^{1}-[p]_-^+
\hat{\mathbf{n}}\partial_{\alpha} s\,\d \alpha,
\end{equation}
where $\mathbf{F}_\mathrm{LES}(t)$ represents the force generated from the leading edge suction, and $[p](\alpha,t)$ is obtained by integrating~\eqref{eq:pressure} and using the boundary condition $[p](1,t)=0$.
The leading-edge suction force used is the limit of the suction force on a leading-edge of small but finite radius of curvature, in the limit that the radius tends to zero. This is a reasonable model of the actual flow when it is attached~\cite{saffman1992vortex}, so we choose to include it. At the membrane's leading edge $\alpha=-1$, $\gamma=v/\sqrt{1-\alpha^2}$ has a singularity which requires an external force to be applied at the tip of the membrane; we call this the leading edge suction.
Following~\cite[pp.\ 97--98]{saffman1992vortex}, we derive the force due to leading edge suction as follows. First, we obtain the velocity components $(u,v)$ through 
\begin{align}
u-iv&=\frac{\d w^{\pm}}{\d z}\sim \mp \frac{\gamma}{2}e^{-i\theta_{-1}}\sim\mp\frac{v_{-1}e^{-i\theta_{-1}}}{2\sqrt{1-\alpha^2}}\sim\mp\frac{v_{-1}e^{-i\theta_{-1}}}{2\sqrt{2}\sqrt{1+\alpha}}\sim \mp\frac{v_{-1}e^{-i\theta_{-1}}}{2\sqrt{2}\sqrt{\alpha-\alpha_{\mathrm{LE}}}}\sim\mp\frac{v_{-1}e^{-i\theta_{-1}}}{2\sqrt{2}\sqrt{z}},
\end{align}
where the subscript $-1$ is used to denote function evaluation at the leading edge material coordinate $\alpha=-1$ (i.e., $v_{-1}=v(-1,t)$ is the membrane's velocity at its leading edge).
If we assume that $w=cz^{1/2}$ then 
\begin{align}
\frac{\d w}{\d z}=\frac{c}{2}\frac{1}{\sqrt{z}}=-\frac{v_{-1}e^{-i\theta_{-1}}}{2\sqrt{2}}\frac{1}{\sqrt{z}},
\end{align}
where $c=-v_{-1}e^{-i\theta_{-1}}/\sqrt{2}$.
If one then computes the leading edge suction force as the pressure force on a small circle centered at the leading edge as in~\cite[pp.\ 97--98]{saffman1992vortex}, one obtains
\begin{equation}\label{eq:FLES}
\mathbf{F}_{\mathrm{LES}}(t)=-\frac{\pi}{8}v_{-1}^2\,\hat{\mathbf{s}}_{-1},
\end{equation}
which is in accord with~\cite[eq.\ (2.23)]{alben2008optimal} and~\cite[eq.\ (B.7)]{moore2017fast}.
Similarly to~\cite{alben2009simulating} we use the transformation $\alpha=-\cos\varphi$, where $\varphi\in[0,\pi]$ such that $\alpha\in[-1,1]$, to write~\eqref{eq:force1} as
\begin{equation}\label{eq:ftot}
\mathbf{F}(t)=\mathbf{F}_\mathrm{LES}(t)+\int_{0}^{\pi}-[p]_-^+
\hat{\mathbf{n}}\partial_{\alpha} s\sin\varphi\,\d \varphi,
\end{equation}
where $\sin\varphi=\sqrt{1-\alpha^2}$. The pressure jump has a square root singularity at the leading edge but $[p]\sin\varphi$ is bounded.

The drag force $\mathrm{D}(t)$ and lift force $\mathrm{L}(t)$  are thus given by the horizontal and vertical components of~\eqref{eq:ftot} respectively:
\begin{equation}
    (\mathrm{D}(t),\mathrm{L}(t)) = \mathbf{F}(t).
\end{equation}
 
By D'Alembert's paradox, the drag force on a body in steady potential flow is zero, and this holds to a good approximation for our numerical solutions.




\bibliographystyle{unsrt}
\bibliography{biblio.bib}

\begin{thebibliography}{10}

\bibitem{nielsen1963theory}
J.~N. Nielsen.
\newblock Theory of flexible aerodynamic surfaces.
\newblock {\em J. Applied Mech.}, 30(3):435--442, 1963.

\bibitem{newman1984two}
B.~G. Newman and H.~T. Low.
\newblock Two-dimensional impervious sails: experimental results compared with
  theory.
\newblock {\em J. Fluid Mech.}, 144:445--462, 1984.

\bibitem{newman1987aerodynamic}
B.~G. Newman.
\newblock Aerodynamic theory for membranes and sails.
\newblock {\em Progress in Aerospace Sciences}, 24(1):1--27, 1987.

\bibitem{smith1995computation}
R.~Smith and W.~Shyy.
\newblock Computation of unsteady laminar flow over a flexible two-dimensional
  membrane wing.
\newblock {\em Physics of Fluids}, 7(9):2175--2184, 1995.

\bibitem{vanden1981shape}
J.-M. Vanden-Broeck and J.~B. Keller.
\newblock Shape of a sail in a flow.
\newblock {\em Physics of Fluids}, 24(3):552--553, 1981.

\bibitem{vanden1982nonlinear}
J.-M. Vanden-Broeck.
\newblock Nonlinear two-dimensional sail theory.
\newblock {\em The Physics of Fluids}, 25(3):420--423, 1982.

\bibitem{kimball2009physics}
J.~Kimball.
\newblock {\em Physics of sailing}.
\newblock CRC Press, Boca Raton, FL, 2009.

\bibitem{pepper1971aerodynamic}
W.~B. Pepper and R.~C. Maydew.
\newblock Aerodynamic decelerators-an engineering review.
\newblock {\em Journal of Aircraft}, 8(1):3--19, 1971.

\bibitem{stein2000parachute}
K.~Stein, R.~Benney, V.~Kalro, T.~E. Tezduyar, J.~Leonard, and M.~Accorsi.
\newblock Parachute fluid--structure interactions: {3-D} computation.
\newblock {\em Computer Methods in Applied Mechanics and Engineering},
  190(3-4):373--386, 2000.

\bibitem{shyy1999flapping}
W.~Shyy, M.~Berg, and D.~Ljungqvist.
\newblock Flapping and flexible wings for biological and micro air vehicles.
\newblock {\em Progress in aerospace sciences}, 35(5):455--505, 1999.

\bibitem{lian2003membrane}
Y.~Lian, W.~Shyy, D.~Viieru, and B.~Zhang.
\newblock Membrane wing aerodynamics for micro air vehicles.
\newblock {\em Progress in Aerospace Sciences}, 39(6-7):425--465, 2003.

\bibitem{albertani2007aerodynamic}
R.~Albertani, B.~Stanford, J.~P. Hubner, and P.~G. Ifju.
\newblock Aerodynamic coefficients and deformation measurements on flexible
  micro air vehicle wings.
\newblock {\em Experimental Mechanics}, 47:625--635, 2007.

\bibitem{hu2008flexible}
H.~Hu, M.~Tamai, and J.~T. Murphy.
\newblock Flexible-membrane airfoils at low {R}eynolds numbers.
\newblock {\em Journal of Aircraft}, 45(5):1767--1778, 2008.

\bibitem{stanford2008fixed}
B.~Stanford, P.~Ifju, R.~Albertani, and W.~Shyy.
\newblock Fixed membrane wings for micro air vehicles: Experimental
  characterization, numerical modeling, and tailoring.
\newblock {\em Progress in Aerospace Sciences}, 44(4):258--294, 2008.

\bibitem{scott2007aeroelastic}
R.~Scott, R.~Bartels, and O.~Kandil.
\newblock An aeroelastic analysis of a thin flexible membrane.
\newblock In {\em 48th AIAA/ASME/ASCE/AHS/ASC Structures, Structural Dynamics,
  and Materials Conference}, page 2316, 2007.

\bibitem{rohrschneider2007survey}
R.~R. Rohrschneider and R.~D. Braun.
\newblock Survey of ballute technology for aerocapture.
\newblock {\em Journal of Spacecraft and Rockets}, 44(1):10--23, 2007.

\bibitem{voss1961effect}
H.~M. Voss.
\newblock The effect of an external supersonic flow on the vibration
  characteristics of thin cylindrical shells.
\newblock {\em Journal of the Aerospace Sciences}, 28(12):945--956, 1961.

\bibitem{ashley1956piston}
H.~Ashley and G.~Zartarian.
\newblock Piston theory-a new aerodynamic tool for the aeroelastician.
\newblock {\em Journal of the aeronautical sciences}, 23(12):1109--1118, 1956.

\bibitem{haruo1975flutter}
K.~Haruo.
\newblock Flutter of hanging roofs and curved membrane roofs.
\newblock {\em International Journal of Solids and Structures}, 11(4):477--492,
  1975.

\bibitem{knudson1991recent}
W.~C. Knudson.
\newblock Recent advances in the field of long span tension structures.
\newblock {\em Engineering Structures}, 13(2):164--177, 1991.

\bibitem{sygulski1996dynamic}
R.~Sygulski.
\newblock Dynamic stability of pneumatic structures in wind: theory and
  experiment.
\newblock {\em J. Fluids and Struct.}, 10(8):945--963, 1996.

\bibitem{sygulski1997numerical}
R.~Sygulski.
\newblock Numerical analysis of membrane stability in air flow.
\newblock {\em Journal of Sound and Vibration}, 201(3):281--292, 1997.

\bibitem{sygulski2007stability}
R.~Sygulski.
\newblock Stability of membrane in low subsonic flow.
\newblock {\em Inter. J. of Non-Lin. Mech.}, 42(1):196--202, 2007.

\bibitem{swartz1996mechanical}
S.~M. Swartz, M.~S. Groves, H.~D. Kim, and W.~R. Walsh.
\newblock Mechanical properties of bat wing membrane skin.
\newblock {\em Journal of Zoology}, 239(2):357--378, 1996.

\bibitem{song2008aeromechanics}
A.~Song, X.~Tian, E.~Israeli, R.~Galvao, K.~Bishop, S.~Swartz, and K.~Breuer.
\newblock Aeromechanics of membrane wings with implications for animal flight.
\newblock {\em AIAA journal}, 46(8):2096, 2008.

\bibitem{cheney2015wrinkle}
J.~A. Cheney, N.~Konow, A.~Bearnot, and S.~M. Swartz.
\newblock A wrinkle in flight: the role of elastin fibres in the mechanical
  behaviour of bat wing membranes.
\newblock {\em Journal of the Royal Society Interface}, 12(106):20141286, 2015.

\bibitem{anderson2008physics}
B.~D. Anderson.
\newblock The physics of sailing.
\newblock {\em Physics Today}, 61(2):38--43, 2008.

\bibitem{maria2013recent}
I.~M. Viola.
\newblock Recent advances in sailing yacht aerodynamics.
\newblock {\em Applied Mechanics Reviews}, 65(4):040801, 2013.

\bibitem{lombardi2012strongly}
M.~Lombardi, M.~Cremonesi, A.~Giampieri, N.~Parolini, A.~Quarteroni, et~al.
\newblock A strongly coupled fluid-structure interaction model for wind-sail
  simulation.
\newblock {\em 4th High Performance Yacht Design}, pages 212--221, 2012.

\bibitem{petres2012potential}
C.~P{\^e}tr{\`e}s, M.-A. Romero-Ramirez, and F.~Plumet.
\newblock A potential field approach for reactive navigation of autonomous
  sailboats.
\newblock {\em Robotics and Autonomous Systems}, 60(12):1520--1527, 2012.

\bibitem{mavroyiakoumou2020large}
C.~Mavroyiakoumou and S.~Alben.
\newblock Large-amplitude membrane flutter in inviscid flow.
\newblock {\em Journal of Fluid Mechanics}, 891:A23, 2020.

\bibitem{souppez2019recent}
J.-B.~R.~G. Souppez, A.~Arredondo-Galeana, and I.~M. Viola.
\newblock Recent advances in numerical and experimental downwind sail
  aerodynamics.
\newblock {\em Journal of Sailing Technology}, 4(01):45--65, 2019.

\bibitem{arredondo2023vortex}
A.~Arredondo-Galeana, H.~Babinsky, and I.~M. Viola.
\newblock Vortex flow of downwind sails.
\newblock {\em Flow}, 3:E8, 2023.

\bibitem{greenhalgh1984aerodynamic}
S.~Greenhalgh, H.~C. Curtiss~Jr, and B.~Smith.
\newblock Aerodynamic properties of a two-dimensional inextensible flexible
  airfoil.
\newblock {\em AIAA journal}, 22(7):865--870, 1984.

\bibitem{hess1961experimental}
R.~W. Hess.
\newblock {\em An Experimental Study of the Flutter of Sails Having a Delta
  Planform Tested from a Mach Number of 0.1 to a Mach Number of 1.9}.
\newblock National Aeronautics and Space Administration, 1961.

\bibitem{windsportatlanta}
WindsportAtlanta.com.
\newblock {Lowest batten doesn't pop into right shape for wind direction}.
\newblock
  \url{https://windsportatlanta.com/content/lowest-batten-doesnt-pop-right-shape-wind-direction-11918-0}.
\newblock 2018, accessed online 22 July 2024.

\bibitem{cruisersforum}
cruisersforum.com.
\newblock {S-Shape in Main going upwind. Why?}
\newblock
  \url{https://www.cruisersforum.com/forums/f48/s-shape-in-main-going-upwind-why-170685-2.html}.
\newblock 2016, accessed online 22 July 2024.

\bibitem{catsailor}
Catamaran sailor.
\newblock {S-Bend in the mainsail}.
\newblock \url{https://www.catsailor.com/forum/ubbthreads.php/topics/38354/1}.
\newblock 2001, accessed online 22 July 2024.

\bibitem{kim2021flow}
H.~Kim, M.~Lahooti, J.~Kim, and D.~Kim.
\newblock Flow-induced periodic snap-through dynamics.
\newblock {\em Journal of Fluid Mechanics}, 913:A52, 2021.

\bibitem{kim2021snap}
J.~Kim, H.~Kim, and D.~Kim.
\newblock Snap-through oscillations of tandem elastic sheets in uniform flow.
\newblock {\em Journal of Fluids and Structures}, 103:103283, 2021.

\bibitem{chen2023snap}
Z.~Chen, Q.~Mao, Y.~Liu, and H.~J. Sung.
\newblock Snap-through dynamics of a buckled flexible filament with different
  edge conditions.
\newblock {\em Physics of Fluids}, 35(10), 2023.

\bibitem{shoele2016energy}
K.~Shoele and R.~Mittal.
\newblock Energy harvesting by flow-induced flutter in a simple model of an
  inverted piezoelectric flag.
\newblock {\em Journal of Fluid Mechanics}, 790:582--606, 2016.

\bibitem{orrego2017harvesting}
S.~Orrego, K.~Shoele, A.~Ruas, K.~Doran, B.~Caggiano, R.~Mittal, and S.~H.
  Kang.
\newblock Harvesting ambient wind energy with an inverted piezoelectric flag.
\newblock {\em Applied energy}, 194:212--222, 2017.

\bibitem{kim2013flapping}
D.~Kim, J.~Coss{\'e}, C.~H. Cerdeira, and M.~Gharib.
\newblock Flapping dynamics of an inverted flag.
\newblock {\em J. Fluid Mech.}, 736, 2013.

\bibitem{ryu2015flapping}
J.~Ryu, S.~G. Park, B.~Kim, and H.~J. Sung.
\newblock Flapping dynamics of an inverted flag in a uniform flow.
\newblock {\em Journal of Fluids and Structures}, 57:159--169, 2015.

\bibitem{gurugubelli2015self}
P.~S. Gurugubelli and R.~K. Jaiman.
\newblock Self-induced flapping dynamics of a flexible inverted foil in a
  uniform flow.
\newblock {\em Journal of Fluid Mechanics}, 781:657--694, 2015.

\bibitem{gomez2017passive}
M.~Gomez, D.~E. Moulton, and D.~Vella.
\newblock Passive control of viscous flow via elastic snap-through.
\newblock {\em Physical review letters}, 119(14):144502, 2017.

\bibitem{jouffroy2009control}
J.~Jouffroy.
\newblock A control strategy for steering an autonomous surface sailing vehicle
  in a tacking maneuver.
\newblock In {\em 2009 IEEE International Conference on Systems, Man and
  Cybernetics}, pages 2391--2396. IEEE, 2009.

\bibitem{sanchez2020autonomous}
P.~J.~B. S{\'a}nchez, M.~Papaelias, and F.~P.~G. M{\'a}rquez.
\newblock Autonomous underwater vehicles: Instrumentation and measurements.
\newblock {\em IEEE Instrumentation \& Measurement Magazine}, 23(2):105--114,
  2020.

\bibitem{bandyopadhyay2005trends}
P.~R. Bandyopadhyay.
\newblock Trends in biorobotic autonomous undersea vehicles.
\newblock {\em IEEE Journal of Oceanic Engineering}, 30(1):109--139, 2005.

\bibitem{tranzatto2015debut}
M.~Tranzatto, A.~Liniger, S.~Grammatico, and A.~Landi.
\newblock {The debut of Aeolus, the autonomous model sailboat of ETH Zurich}.
\newblock In {\em OCEANS 2015-Genova}, pages 1--6. IEEE, 2015.

\bibitem{tranzatto2015navigation}
M.~Tranzatto.
\newblock {\em Navigation and Control for an Autonomous Sailing Model Boat}.
\newblock PhD thesis, University of Pisa, 2015.

\bibitem{stelzer2010reactive}
R.~Stelzer, K.~Jafarmadar, H.~Hassler, and R.~Charwot.
\newblock A reactive approach to obstacle avoidance in autonomous sailing.
\newblock In {\em 3rd International Robotic Sailing Conference (IRSC 2010),
  Kingston, Ontario, Canada}, pages 34--40, 2010.

\bibitem{xiao2012wind}
K.~Xiao, J.~Sliwka, and L.~Jaulin.
\newblock A wind-independent control strategy for autonomous sailboats based on
  voronoi diagram.
\newblock In {\em Field Robotics}, pages 110--124. World Scientific, 2012.

\bibitem{tipsuwan2023overview}
Y.~Tipsuwan, P.~Sanposh, and N.~Techajaroonjit.
\newblock Overview and control strategies of autonomous sailboats—a survey.
\newblock {\em Ocean Engineering}, 281:114879, 2023.

\bibitem{hansen2019maneuver}
H.~Hansen, K.~Hochkirch, I.~Burns, and S.~Ferguson.
\newblock Maneuver simulation and optimization for {AC50} class.
\newblock {\em Journal of Sailing Technology}, 4(01):142--160, 2019.

\bibitem{waldman2013shape}
R.~M. Waldman and K.~S. Breuer.
\newblock Shape, lift, and vibrations of highly compliant membrane wings.
\newblock In {\em 43rd AIAA Fluid Dynamics Conference}, page 3177, 2013.

\bibitem{gerhardt2011unsteady}
F.~C. Gerhardt, R.~G.~J. Flay, and P.~Richards.
\newblock Unsteady aerodynamics of two interacting yacht sails in
  two-dimensional potential flow.
\newblock {\em Journal of Fluid Mechanics}, 668:551--581, 2011.

\bibitem{young2019effect}
J.~D. Young, S.~E. Morris, R.~R. Schutt, and C.~H.~K. Williamson.
\newblock Effect of hybrid-heave motions on the propulsive performance of an
  oscillating airfoil.
\newblock {\em Journal of Fluids and Structures}, 89:203--218, 2019.

\bibitem{schutt2016unsteady}
R.~R. Schutt and C.~H.~K. Williamson.
\newblock Unsteady sail dynamics due to bodyweight motions.
\newblock In {\em Chesapeake Sailing Yacht Symposium}. SNAME, 2016.

\bibitem{mavroyiakoumou2021dynamics}
C.~Mavroyiakoumou and S.~Alben.
\newblock Dynamics of tethered membranes in inviscid flow.
\newblock {\em Journal of Fluids and Structures}, 107:103384, 2021.

\bibitem{carrier1945non}
G.~F. Carrier.
\newblock On the non-linear vibration problem of the elastic string.
\newblock {\em Quarterly of Applied Mathematics}, 3(2):157--165, 1945.

\bibitem{narasimha1968non}
R.~Narasimha.
\newblock Non-linear vibration of an elastic string.
\newblock {\em Journal of Sound and Vibration}, 8(1):134--146, 1968.

\bibitem{nayfeh2008linear}
A.~H. Nayfeh and P.~F. Pai.
\newblock {\em Linear and nonlinear structural mechanics}.
\newblock John Wiley \& Sons, New York, 2008.

\bibitem{saffman1992vortex}
P.~G. Saffman.
\newblock {\em Vortex dynamics}.
\newblock Cambridge University Press, 1992.

\bibitem{reinsch1967smoothing}
C.~H. Reinsch.
\newblock Smoothing by spline functions.
\newblock {\em Numerische mathematik}, 10(3):177--183, 1967.

\bibitem{MATLAB}
The~MathWorks Inc.
\newblock Matlab version: R2023a, 2023.

\bibitem{irvine1979note}
H.~M. Irvine.
\newblock A note on luffing in sails.
\newblock {\em Proceedings of the Royal Society of London. A. Mathematical and
  Physical Sciences}, 365(1722):345--347, 1979.

\bibitem{tiomkin2017stability}
S.~Tiomkin and D.~E. Raveh.
\newblock On the stability of two-dimensional membrane wings.
\newblock {\em J. Fluids and Struct.}, 71:143--163, 2017.

\bibitem{newman1991stability}
B.~G. Newman and M.~P. Pa\"idoussis.
\newblock The stability of two-dimensional membranes in streaming flow.
\newblock {\em J. Fluids and Struct.}, 5(4):443--454, 1991.

\bibitem{le1999unsteady}
O.~Le~Ma{\^\i}tre, S.~Huberson, and E.~S. De~Cursi.
\newblock Unsteady model of sail and flow interaction.
\newblock {\em J. Fluids and Struct.}, 13(1):37--59, 1999.

\bibitem{mavroyiakoumou2022membrane}
C.~Mavroyiakoumou and S.~Alben.
\newblock Membrane flutter in three-dimensional inviscid flow.
\newblock {\em Journal of Fluid Mechanics}, 953:A32, 2022.

\bibitem{mavroyiakoumou2023spanwise}
C.~Mavroyiakoumou and S.~Alben.
\newblock Spanwise variations in membrane flutter dynamics.
\newblock {\em arXiv preprint arXiv:2311.16443}, 2023.

\bibitem{northsails}
North Sails.
\newblock {How to improve your tacks: Prioritizing Your Upwind Technique Can
  Make Or Break Your Race Result}.
\newblock
  \url{https://www.northsails.com/en-it/blogs/north-sails-blog/how-to-improve-your-tacks-north-sails
  }.
\newblock 5 August, accessed online 8 September 2024.

\bibitem{bethwaite2013fast}
F.~Bethwaite.
\newblock {\em Fast handling technique}.
\newblock A\&C Black, Bloomsbury Publishing, 2013.

\bibitem{sneyd1984aerodynamic}
A.~D. Sneyd.
\newblock Aerodynamic coefficients and longitudinal stability of sail
  aerofoils.
\newblock {\em Journal of Fluid Mechanics}, 149:127--146, 1984.

\bibitem{batchelor1967introduction}
G.~K. Batchelor.
\newblock {\em An introduction to fluid dynamics}.
\newblock Cambridge {U}niversity {P}ress, 1967.

\bibitem{acheson1990elementary}
D.~J. Acheson.
\newblock {\em Elementary fluid dynamics}.
\newblock Oxford University Press, 1990.

\bibitem{alben2008optimal}
S.~Alben.
\newblock Optimal flexibility of a flapping appendage in an inviscid fluid.
\newblock {\em J. Fluid Mech.}, 614:355--380, 2008.

\bibitem{moore2017fast}
M.~N.~J. Moore.
\newblock A fast {C}hebyshev method for simulating flexible-wing propulsion.
\newblock {\em Journal of Computational Physics}, 345:792--817, 2017.

\bibitem{alben2009simulating}
S.~Alben.
\newblock Simulating the dynamics of flexible bodies and vortex sheets.
\newblock {\em Journal of Computational Physics}, 228(7):2587--2603, 2009.

\end{thebibliography}

\end{document}